\documentclass[preprint,preprintnumbers,amsmath,amssymb]{revtex4-1}

\usepackage{graphicx}% Include figure files
\usepackage{dcolumn}% Align table columns on decimal point
\usepackage{bm}% bold math
\usepackage[dvips]{color}
\usepackage{ulem}

\begin{document}

\preprint{APS/123-QED}

\title{Long-range interactions in the ozone molecule: spectroscopic and
dynamical points of view}

\author{Maxence Lepers$^1$, B\'eatrice Bussery-Honvault$^2$ and Olivier Dulieu$^1$}
\affiliation{$^1$Laboratoire Aim\'e Cotton, CNRS/Univ. Paris-Sud/ENS-Cachan, Orsay, France \\
$^2$Laboratoire Interdisciplinaire Carnot de Bourgogne, CNRS--Universit\'e de Bourgogne, Dijon, France}
\email{maxence.lepers@u-psud.fr}

\date{\today}

\begin{abstract}
Using the multipolar expansion of the electrostatic energy, we have characterized the asymptotic interactions between an oxygen atom O$(^3P)$ and an oxygen molecule O$_2(^3\Sigma_g^-)$, both in their electronic ground state. We have calculated the interaction energy induced by the permanent electric quadrupoles of O and O$_2$ and the van der Waals energy. On one hand we determined the 27 electronic potential energy surfaces including spin-orbit connected to the O$(^3P)$ + O$_2(^3\Sigma_g^-)$ dissociation limit of the O--O$_2$ complex. On the other hand we computed the potential energy curves characterizing the interaction between O$(^3P)$ and a O$_2(^3\Sigma_g^-)$ molecule in its lowest vibrational level and in a low rotational level. Such curves are found adiabatic to a good approximation, namely they are only weakly coupled to each other. These results represent a first step for modeling the spectroscopy of ozone bound levels close to the dissociation limit, as well as the low energy collisions between O and O$_2$ thus complementing the knowledge relevant for the ozone formation mechanism.

%Firstly, we have considered the O--O$_2$ complex in fixed geometries, \textit{i.e.}~with given distances and angles, and so we have calculated the 27 spin-orbit potential energy surfaces connected to the O$(^3P)$ + O$_2(^3\Sigma_g^-)$ dissociation limit. Secondly, we have considered the situation where O$_2$ vibrates in the ground level and asymptotically rotates in an arbitrary level. We have obtained potential energy curves functions of the O--O$_2$ distance $R$ which are found to be adiabatic to a good approximation, \textit{i.e.}~weakly coupled with each other. Our results are relevant for the spectroscopy of the O--O$_2$ complex close to its dissociation limit, and for the low energy collision of O and O$_2$ in the asymptotic region, which may complete the knowledge on the ozone formation mechanism.
\end{abstract}

%\pacs{Valid PACS appear here}% PACS, the Physics and Astronomy
                             % Classification Scheme.
%\keywords{Suggested keywords}%Use showkeys class option if keyword
                              %display desired
\maketitle

\section{Introduction}

The ozone molecule plays a crucial role in the physics and chemistry of the Earth atmosphere. However, a lot remains to be understood, especially about its formation, which is thought to take place in two steps \cite{schinke2006}. Firstly, an oxygen atom O and an oxygen molecule O$_2$ collide to give a ro-vibrationally or electronically excited ozone complex O$_3^*$. Secondly, this complex stabilizes by inelastic collision with one surrounding atom or molecule, which is the so-called \textit{deactivation} process. However, this second step takes place provided that the excited complex O$_3^*$ does not dissociate into O$+$O$_2$ before colliding with the surrounding gas. Characterizing the competition between deactivation on the one hand, and dissociation on the other hand, is the key point in order to understand quantitatively the ozone formation in the atmosphere.

In this respect, one of the most striking features of ozone physical chemistry is the unconventional isotopic effects that influence the competition between stabilization and dissociation of O$_3^*$ \cite{thiemens1983,mauersberger1987,anderson1997,janssen2001,mauersberger2005}. It became clear in recent years \cite{janssen1999}, that these isotopic effects were determined by the difference of zero-point energies  $\Delta\textrm{ZPE}$ of the O$_2$ isotopologues in the entrance and the dissociation channels. If $\Delta\textrm{ZPE}>0$, dissociation is energetically unfavorable: the O$_3^*$ has a higher lifetime, and so is more likely to give stable O$_3$. On the contrary, if $\Delta\textrm{ZPE}<0$, dissociation is energetically favorable, and it tends to dominate stabilization.

The unconventional isotopic effects were very well understood in the beginning of the 2000's within the framework of the statistical RKRM (Rice-Kassel-Ramsperger-Marcus) theory \cite{gao2001,gao2002}. However, an adjustable parameter $\eta$ had to be added to that theory, in order to account for deviation from the energy equipartition theorem, after the formation of the O$_3^*$ molecule. Then, the need for first-principle studies of the ozone formation, especially based on quantum mechanics, became obvious. Since a full quantum treatment of the two-step process is beyond the possibilities offered by current computers, researchers are urged to focus on precise aspects of the process, \textit{e.g.} highly-excited vibrational levels of O$_3$ \cite{grebenshchikov2003,babikov2003,lee2004}, the influence of resonances \cite{babikov2003b,grebenshchikov2009}, or to use less demanding computational techniques, \textit{e.g.} classical-trajectory \cite{fleurat2003} or mixed-quantum-classical calculations \cite{ivanov2012}.

The common features of all those studies is that they need a reliable potential energy surface (PES), at least for the electronic ground state $\tilde{X}^1A_1$. Since the formation of stable O$_3$ involves a wide variety of geometries, from almost separated O and O$_2$, to tightly bound O$_3$, one actually needs a \textit{global} PES. Up to now, all the published PESs share the same general features. In particular when they are cut along the minimum-energy path, they all show a change in character between the inner and the asymptotic regions, which is due to an avoided crossing with an excited electronic state, also referred to as the transition state. However, there is still a controversy on whether this avoided crossing induces a potential barrier that goes above \cite{siebert2001,siebert2002} or below the dissociation limit \cite{rosmus2002,babikov2003b,holka2010}, or on the contrary, a monotonic evolution of the potential energy as suggested by the most recent \textit{ab initio} calculations \cite{dawes2011}.

Up to now, all the articles that aimed at describing the O$-$O$_2$ asymptotic region were based on quantum-chemical calculations of the ground-state and possibly the lowest excited-states PESs of O$_3$ \cite{rosmus2002,tashiro2003}. In the present paper, we propose an alternative method based on the multipolar expansion of the electrostatic potential energy between O and O$_2$, both in their electronic ground state. This method enables us to obtain, in a single calculation, the 27 spin-orbit PESs connected to the dissociation threshold O$(^3P)+$O$_2(X^3\Sigma_g^-)$. The calculated PESs are then functions of the electric properties, \textit{i.e.} multipole moments and multipole polarizabilities, of the separated systems. This method is valid provided that the electronic clouds of O and O$_2$ do not overlap, that is for a O--O$_2$ distance larger than 8 Bohr. Such asymptotic PESs can then be used as a tool to check the quality of the \textit{ab initio} PESs.

In this article, we use two complementary approaches to characterize the O$-$O$_2$ long-range interactions. In section \ref{sec:FixedGeom}, we consider the atom and the diatom at \textit{fixed} geometries, that is at given inter-particle distances and bending angles. The obtained PESs can be directly connected to \textit{ab initio} PESs. In section \ref{sec:O2-rovib}, we include in our model the vibration and rotation of O$_2$ which yields one-dimensional potential energy curves (PECs) depending on the O--O$_2$ distance. Such curves are relevant for the low energy dynamics of the O--O$_2$ complex, in particular to which extent the rotation of O$_2$ is hindered by the presence of O. These curves are also useful for modeling the vibrational levels of ozone close to its dissociation limit \cite{grebenshchikov2003, babikov2003, lee2004}. We discuss in section \ref{sec:Discussion} the way to connect both approaches which is done for the first time, to the best of our knowledge. Section \ref{sec:Conclusion} contains our conclusions and prospects.

\section{\label{sec:FixedGeom}Asymptotic electronic potential energy surfaces between {O} and {O}$_2$}

As the two ground state particles
O and O$_{2}$ are far away from each other, \textit{i.e.}
their electronic clouds do not overlap, we use the Jacobi coordinates
to describe the PESs. The body-fixed frame associated to the O--O$_2$ complex has its $z$ axis connecting O to the center of mass of
O$_{2}$. The $x$ axis is perpendicular to the $z$ axis and is located
in the plane of the three atoms; the $y$ axis is perpendicular to
this plane. We also introduce the coordinate system $x_{d}y_{d}z_{d}$
linked to the O$_{2}$ diatom. The axes $z_{d}$ and $x_{d}$ are related
to $z$ and $x$ by a rotation of angle $\theta$ about the $y_{d}\equiv y$
axis. We denote as $r$ the interatomic distance in the O$_{2}$ molecule,
and $R$ the distance between the O atom and the center of mass of
O$_{2}$. The formalism presented in this section allows for
calculating three-dimensional PESs depending on $R$, $r$ and $\theta$.
In the following, we set $r \equiv r_{e}=2.282a_0$ ($a_0=0.0529177$~nm is the Bohr radius and the atomic unit for distances) at the equilibrium distance
of O$_{2}$ without loss of generality for our purpose.

\subsection{Model}

The calculations below are based on the formalism
described in Ref.~\cite{bussery-honvault2008}, and used in Ref.~\cite{bussery-honvault2009}.
The quantities referring to O and O$_{2}$ are respectively characterized
by the subscripts {}``$a$'' after {}``atom'' and {}``$d$''
after {}``diatom''. The starting point of our model is the multipole
expansion in inverse powers of $R$ of the electrostatic energy
between two charge distributions {}``$a$'' and {}``$d$'' (see
\cite{bussery-honvault2008}, Eqs.~(2) and (3))
\begin{equation}
\hat{V}_{ad}=\sum_{\ell_{a}\ell_{d}}\frac{1}{R^{1+\ell_{a}+\ell_{d}}}\sum_{m=-\ell_{<}}^{+\ell_{<}} g_{m}(\ell_{a},\ell_{d})\hat{Q}_{\ell_{a}}^{m}\hat{Q}_{\ell_{d}}^{-m}
\label{eq:Vad}
\end{equation}
with $\ell_{<}=\min(\ell_{a},\ell_{d})$ and  \cite{lepers2010}
\begin{equation}
g_{m}(\ell_{a},\ell_{d})=\frac{(-1)^{\ell_{d}}(\ell_{a}+\ell_{d})!}{\sqrt{(\ell_{a}+m)!(\ell_{a}-m)!(\ell_{d}+m)!(\ell_{d}-m)!}}.
\label{eq:g-m}
\end{equation}
In Eq.~(\ref{eq:Vad}), $\hat{Q}_{\ell_{a}}^{m}$ ($\hat{Q}_{\ell_{d}}^{-m}$) is the $\ell_{a}$- ($\ell_{d}$-) rank and $m$-($-m$-)component multipole-moment operator of the atom (diatom) expressed in the $xyz$ coordinate system. The multipole moments of the diatom is conveniently expressed as
\begin{equation}
\hat{Q}_{\ell_{d}}^{-m}=\sum_{m_{d}=-\ell_{d}}^{+\ell_{d}} d_{-mm_{d}}^{\, \ell_{d}}(\theta) \hat{q}_{\ell_{d}}^{m_{d}} \,,
\label{eq:Vad2}
\end{equation}
where $\hat{q}_{\ell_{d}}^{m_{d}}$ is the multipole moment operator of the diatom expressed in the $x_{d}y_{d}z_{d}$ coordinate system, and $d_{-mm_{d}}^{\, \ell_{d}}(\theta)$ is the reduced Wigner function, which characterizes the rotation from the $x_{d}y_{d}z_{d}$ to the $xyz$ frame.

The multipolar expansion (\ref{eq:Vad}) is valid provided that
the electronic wave functions of O and O$_2$ do not overlap. This condition is fulfilled
for distances $R$ larger than the so-called LeRoy radius \cite{leroy1974} $R_{LR}=2\{\sqrt{\langle r^2\rangle_{\mathrm{O}}}+\sqrt{\langle r^2\rangle_{\mathrm{O}_2}}\} \approx 8a_0$, where the mean squared radius of the electronic wave function in O is $\langle r^2\rangle_{\mathrm{O}}=2.001a_0$. The quantity $\sqrt{\langle r^2\rangle_{\mathrm{O}_2}} \approx r_e/2+\langle r\rangle_{\mathrm{O}}$ is roughly evaluated from the equilibrium distance $r_e=2.282a_0$ of O$_2$ and the mean radius of the electronic wave function in O $\langle r\rangle_{\mathrm{O}}=1.24a_0$.

In this work, we calculate the two leading terms of the $R^{-n}$ expansion for $R>R_{LR}$: the first-order term reflecting the interaction between the permanent quadrupole moments of O and O$_{2}$ ($\ell_{a}=\ell_{d}=2$) scaling as $R^{-5}$, and the second-order (Van der Waals) term  related to the interaction between the induced dipole moments ($\ell_{a}=\ell_{d}=1$) scaling as $R^{-6}$. While generally tedious to obtain from \textit{ab initio} calculations higher-order contributions may also be significant around $R_{LR}$. For instance, assuming that the maximal values of the $C_6$ and $C_8$ coefficients for the O-O$_2$ interaction are half of those for $O_2$-$O_2$  \cite{bartolomei2010}) we obtain $C_6\approx 31$~a.u. (close to the values obtained in the present work, see next section) and $C_8\approx 1291$~a.u.. At $R=8a_0$ the $C_8$ term would then represent at most 64\% of the $C_6$ one and 41\% at $R=10a_0$. Therefore our work provides the essentials of the long-range interaction between O and O$_2$. In order to match the present asymptotic expansions to the \textit{ab initio} calculations, the safest way is to use the $C_5$ and $C_6$ values determined below to fit the long-range part of the \textit{ab initio} PESs beyond $R_{LR}$ and to extract the related higher-order terms. Another possibility would be to directly compute the next term $C_8$, but this beyond the scope of the present paper.

\subsubsection{Zeroth-order energies and state vectors}

%The zeroth-order of the perturbation expansion corresponds to the situation where the two fragments are independent from each other. The unperturbed energy is thus the sum of individual energies, and the corresponding state vectors are products of individual state vectors.

The oxygen atom is in an arbitrary fine-structure level $\left|J_{a}M_{a}\right\rangle $ of its ground
state $^{3}P_{J_{a}M_{a}}$, where $J_{a}$ is the total angular momentum with projection $M_{a}$ on the $z$ axis, resulting from the sum of the projections $M_{L_a}$ and $M_{S_a}$ of the orbital $L_{a}$ and spin $S_{a}$ angular momenta of the atom, respectively. Since the spin-orbit constant $A_O$ of oxygen is negative ($A_O=-79.1$~ cm$^{-1}$), the $^{3}P_{2}$ fine-structure level is the lowest in energy.

The ground electronic state $^{3}\Sigma_{g}^{-}$ of O$_2$ has an orbital angular momentum projection $\Lambda_{d}=0$ on the $z_{d}$ axis and a spin $S_{d}=1$ with projection $\Sigma_{d}$ on $z_{d}$. The fine structure in the O$_2$ energy spectrum induced by the spin-spin interaction writes
\begin{eqnarray}
\left\langle \Lambda_{d}S_{d}\Sigma_{d}\left|\hat{H}_{SS}\right|\Lambda_{d}S_{d}\Sigma_{d}\right\rangle  & = & \left\langle \frac{\lambda}{2}\left(3\hat{S}_{z_{d}}^{2}-\hat{\mathbf{S}}^{2}\right)\right\rangle \nonumber \\
 & = & \frac{\lambda(r)}{2}\left(3\Sigma_{d}^{2}-S_{d}\left(S_{d}+1\right)\right),
\label{eq:Hss}
\end{eqnarray}
where $\lambda(r)$ is the ($r$-dependent) spin-spin constant, with $\lambda(r_{e})=1.980$~cm$^{-1}$ \cite{tinkham1955}. The interaction between the electric multipole moments of O and O$_{2}$ only depends on the spatial coordinates of the electrons and the nuclei so that we will ignore the O$_2$ fine structure in the following. Assuming an energy origin at the $^{3}P_{2}$ oxygen level and at the O$_2$ ground level, the zeroth-order energy is
\begin{equation}
E_{0}=\frac{A_O}{2}\left(J_{a}\left(J_{a}+1\right)-6\right),
\end{equation}
corresponding to the unperturbed state vectors $|\Psi_{0}\rangle=\left|J_{a}M_{a}\right\rangle \left|\Lambda_d\Sigma_{d}\right\rangle $.

\subsubsection{First-order quadrupole-quadrupole interaction}

The first-order quadrupole-quadrupole interaction is obtained by setting $\ell_{a}=\ell_{d}=2$ and $n=5$ in
Eq.~(3) of Ref.~\cite{bussery-honvault2008}. Since $L_a=S_a=1$,  we can rewrite
Eqs.~(4)-(5) of Ref.~\cite{bussery-honvault2008} as
\begin{eqnarray}
 & & E_{\mathrm{elec}}^{\,J_a M_a J'_a M'_a\Lambda_d\Sigma_d\Lambda'_d\Sigma'_d} (r,R,\theta) \nonumber\\
  & = &
\frac{1}{R^{5}} \delta_{\Sigma_d\Sigma'_d}
\sum_{M_{L_a}M_{S_a}} C_{1M_{L_a}1M_{S_a}}^{J_a M_a} \nonumber \\
& \times & \sum_{M'_{L_a}M'_{S_a}} C_{1M'_{L_a}1M'_{S_a}}^{J'_a M'_a} \delta_{M_{S_a}M'_{S_a}} \nonumber \\
& \times & {}^{ii'jj'}V_{52mm_{d}}^{\,\mathrm{elec}}(r)d_{-mm_d}^{2}(\theta),
\label{eq:Eelec}
\end{eqnarray}
where $m_d$ refers to the components of the 2-rank tensor in the diatom frame, $C_{a\alpha b\beta}^{c\gamma}=\langle a\alpha b\beta|c\gamma\rangle$
is the compact expression of Ref.~\cite{varshalovich1988} for the
Clebsch-Gordan coefficients, and where
\begin{equation}
^{ii'jj'}V_{52mm_d}^{\,\mathrm{elec}}(r)=g_{m}(\ell_{a},\ell_{d})\langle 1M_{L_a}|\hat{Q}_{2}^{m}|1M'_{L_a}\rangle\langle\Lambda_{d}|\hat{q}_2^{m_d}(r)|\Lambda'_{d}\rangle\,.
\label{eq:Velec}
\end{equation}
The superscripts $i,i'$ and $j,j'$ designate in a compact way the quantum numbers of O and O$_2$, respectively, \textit{i.e.} $i=\{L_a=1,M_{L_a}\}$, $i'=\{L'_a=1,M'_{L_a}\}$, $j=\{\Lambda_d=0,\Sigma_d\}$ and $j'=\{\Lambda'_d=0,\Sigma'_d\equiv \Sigma_d \}$. As the O$_2$ electronic ground state is of $\Sigma$ symmetry the only non-zero component of the quadrupole operator is $q(r)=\left\langle \Lambda_{d}=0\right|\hat{q}_{2}^{m_d=0}\left|\Lambda_{d}=0\right\rangle$.
In Eq.~(\ref{eq:Velec}) we only keep the quantum numbers which are not fixed: $^{ii'jj'}V_{52mm_{d}}^{\,\mathrm{elec}}$ and $E_{\mathrm{elec}}^{\,J_a M_a J'_a M'_a\Lambda_d\Sigma_d\Lambda'_d\Sigma'_d}$ can be simplified to $^{M_{L_a}M'_{L_a}}V_{52mm_{d}}^{\,\mathrm{elec}}$ and $E_{\mathrm{elec}}^{\,J_a M_a J'_a M'_a}$ respectively.

Using the Wigner-Eckart theorem, we can connect all matrix elements of O quadrupole moment to a single one, say that for which $M_{L_a}=M'_{L_a}=m=0$,
\begin{equation}
\left\langle 1M_{L_a}\right|\hat{Q}_{2}^{m}\left|1M'_{L_a}\right\rangle
 = \left(-1\right)^{M_{L_a}} \sqrt{\frac{15}{2}}
\left(\begin{array}{ccc}
1 & 2 & 1\\
-M_{L_a} & m & M'_{L_a}\end{array}\right)Q\,,
\label{eq:O-quadrup}
\end{equation}
with $Q=\langle 10|\hat{Q}_{2}^{0}|10\rangle$, and where the symbol  $(...)$ is a Wigner 3-j symbol, which imposes $M_{L_a}=M'_{L_a}+m$.

\subsubsection{Second-order dipole-dipole interaction}

The second-order interaction scales as $R^{-6}$ and results from the dispersion
term due to the induced dipole-induced dipole interaction. It is
calculated by setting $\ell_{a}=\ell'_{a}=\ell_{d}=\ell'_{d}=1$ and $n=6$ in
Eqs.~(6)-(10) of Ref.~\cite{bussery-honvault2008}. In this case, the
coupled polarizabilities (Eq.~(7) of Ref.~\cite{bussery-honvault2008}) are tensors which
can be of rank $\mathcal{L}_{a}=0$, 2 and $\mathcal{L}_{d}=0$, 2, for O and O$_{2}$
respectively. Then, the dispersion energy (see Eq.~(8) of Ref.~\cite{bussery-honvault2008})
can be written as a sum of an isotropic ($\mathcal{L}_{d}=0$, \textit{i.e.} $\theta$-independent) and an anisotropic contribution ($\mathcal{L}_d=2$, \textit{i.e.} $\theta$-dependent)
\begin{eqnarray}
E_{\mathrm{disp}}^{\,J_a M_a J'_a M'_a}(r,R,\theta) & = &
 -\frac{1}{R^{6}} \delta_{\Sigma_d\Sigma'_d}
\sum_{M_{L_a}M_{S_a}} C_{1M_{L_a}1M_{S_a}}^{J_a M_a} \nonumber \\
& \times & \sum_{M'_{L_a}M'_{S_a}} C_{1M'_{L_a}1M'_{S_a}}^{J'_a M'_a} \delta_{M_{S_a}M'_{S_a}} \nonumber \\
& \times &
 \left(^{M_{L_a}M'_{L_a}}V_{6000}^{\mathrm{disp}}(r) + \sum_{\mathcal{M}=-\mathcal{L}_{<}}^{\mathcal{L}_{<}}\sum_{\mathcal{M}_{d}=-2}^{2} {}^{M_{L_a}M'_{L_a}}V_{62\mathcal{MM}_{d}}^{\mathrm{disp}}(r) d_{-\mathcal{MM}_{d}}^{2}(\theta)\right),
\label{eq:Edisp}
\end{eqnarray}
with $\mathcal{L}_{<}=\min(\mathcal{L}_a,\mathcal{L}_d=2)\equiv \mathcal{L}_a$. Note that in Eq.~(\ref{eq:Edisp}), the labels for $E_{\mathrm{disp}}$ and $V^{\mathrm{disp}}$ have been simplified in the same way as for the first-order term (see text after Eq.~(\ref{eq:Velec})).

The dispersion coefficients  ${}^{M_{L_a}M'_{L_a}} V_{6\mathcal{L}_d\mathcal{MM}_d}^{\mathrm{disp}}$ (given by Eq.~(9) of Ref.~\cite{bussery-honvault2008}) depend on dynamical polarizabilities at imaginary frequencies. They are conveniently expressed in terms of coupled polarizabilities for O and O$_2$ respectively, following the definitions introduced in Ref.~\cite{spelsberg1993}
\begin{equation}
{}^{M_{L_a}M'_{L_a}}\alpha_{(11)\mathcal{L}_{a}\mathcal{M}}(i\omega) = \sum_{mm'}\langle1m1m'|\mathcal{L}_{a}\mathcal{M}\rangle {}^{M_{L_a}M'_{L_a}}\alpha_{1m1m'}(i\omega),
\end{equation}
\begin{equation}
\alpha_{(11)\mathcal{L}_{d}\mathcal{M}_d}(i\omega;r) = \sum_{qq'}\langle1q1q'|\mathcal{L}_{d}\mathcal{M}_d\rangle \alpha_{1q1q'}(i\omega;r)
\end{equation}
The uncoupled dynamic polarizabilities ${}^{M_{L_{a}}}\alpha_{1010} \equiv ^{M_{L_{a}}}\alpha_{zz}$ for the $M_{L_{a}}$ sublevels of O$(^{3}P)$  assuming an electric field polarized in the $z$ direction
\begin{equation}
{}^{M_{L_a}}\alpha_{zz}(i\omega) = \frac{\sum_{k=0}^{N-1}({}^{M_{L_a}}a_k) \times  (i\omega)^{2k}}{1+\sum_{k=1}^{N}({}^{M_{L_a}}b_k) \times (i\omega)^{2k}},
\label{eq:O-pola-pade}
\end{equation}
have been calculated by one of us using [N,N-1] Pad\'e approximants $({}^{M_{L_a}}a_k)$ and $({}^{M_{L_a}}b_k)$ \cite{langhoff1970} reported in the supplementary material \cite{epaps} for convenience, and published elsewhere \cite{stoecklin2012}. For O($^3P$) we obtain
\begin{eqnarray}
^{M_{L_a}M'_{L_a}}\alpha_{(11)00}(i\omega) & = & -\frac{^{M_{L_{a}}=0}\alpha_{zz}(i\omega)+2{}^{M_{L_{a}}=\pm1}\alpha_{zz}(i\omega)}{\sqrt{3}}\label{eq:O-pola-coupl-0}\\
^{M_{L_a}M'_{L_a}}\alpha_{(11)2\mathcal{M}}(i\omega) & = & \left(-1\right)^{1-M_{L_a}}\sqrt{5}  \left(\begin{array}{ccc}
1 & 2 & 1 \\
-M_{L_a} & \mathcal{M} & M'_{L_a}\end{array}\right) \nonumber \\
 &  & \times\left(^{M_{L_{a}}=\pm1}\alpha_{zz}(i\omega)-{}^{M_{L_{a}}=0}\alpha_{zz}(i\omega)\right) \,.
\label{eq:O-pola-coupl-2}
\end{eqnarray}
Similarly the coupled dynamical polarizabilities of O$_2$ $\alpha_{(11)\mathcal{L}_{d}\mathcal{M}_{d}}$ are related to the ($r$-dependent) uncoupled ones $\alpha_{1010}(i\omega;r)=\alpha^{\parallel}(i\omega;r)$ (the parallel component along $z_d$) and $\alpha_{111-1}(i\omega;r)=-\alpha^{\bot}(i\omega;r)$ (the perpendicular component with respect to $z_d$) according to \cite{bussery-honvault2008}
\begin{eqnarray}
\alpha_{(11)00}(i\omega;r) & = & -\frac{\alpha^{\parallel}(i\omega;r)+2\alpha^{\bot}(i\omega;r)}{\sqrt{3}}\label{eq:O2-pola-coupl-0}\\
\alpha_{(11)2\mathcal{M}_{d}}(i\omega;r) & = & \sqrt{\frac{2}{3}}\delta_{\mathcal{M}_{d}0}\left(\alpha^{\parallel}(i\omega;r)-\alpha^{\bot}(i\omega;r)\right).
\label{eq:O2-pola-coupl-2}
\end{eqnarray}
Note that just like $m_d$ (see Eqs.~(\ref{eq:Eelec}) and (\ref{eq:Velec})), $\mathcal{M}_d$ is zero because O$_2$ is in a $\Sigma$ electronic state.

\subsection{Asymptotic PESs: results and discussions\label{sub:PES-results}}

\begin{table}
\caption{Quadrupole moment $Q$ and static dipole polarizabilities $^{M_{L_a}}\alpha_{zz}(\omega=0)$ (in a.u.) for  O($^3$P). All values are expressed in atomic units.
\label{tab1}}
\vskip 3ex
\begin{ruledtabular}
\begin{tabular}{cccc}
Ref.   & $Q$ & $^{M_{L_a}=0}\alpha_{zz}$ & $^{M_{L_a}=1}\alpha_{zz}$ \\
\hline
This work$^a$            &-0.95&5.64&4.83   \\
Ref.\cite{Gutsev1998}$^b$&-0.95& -  & -  \\
Ref.\cite{Das1998}$^c$   &  -  &5.86&4.94   \\
Ref.\cite{Medved2000}$^d$&-1.02&5.91&4.89   \\
Ref.\cite{Medved2000}$^e$&-1.04&6.08&4.99  \\
\end{tabular}
\end{ruledtabular}
\vskip 0.2cm
$^a$Full valence CASSCF(6e,5o) with aug-cc-pVQZ basis set \\
$^b$CCSD(T) with aug-cc-pV5Z basis set \\
$^c$CCSD(T) with quadruple-$\zeta$ GTO/CGTO basis set~\\
$^d$CASSCF and $^e$CASPT2 both with triple-$\zeta$ GTO/CGTO basis set \\
\end{table}

\begin{table}
\caption{Quadrupole moment $q$, and parallel $\alpha^\parallel(\omega=0)$ and perpendicular $\alpha^\bot(\omega=0)$ static dipole polarizabilities (in a.u.) for O$_2$(X$^3\Sigma^-_g$). All values calculated \textit{ab initio} are given at the equilibrium distance $r_e=2.282$~a.u. and are expressed in atomic units. The value in parenthesis is calculated for for the lowest vibrational level following Ref.~\cite{lawson1997}.
\label{tab1bis}}
\begin{ruledtabular}
\begin{tabular}{cccc}
Method & $q$ & $\alpha^\parallel$ & $\alpha^\perp$ \\ \hline
Ref.~\cite{lawson1997}$^a$     &-0.2530 (-0.2273)          &                &         \\
Ref.~\cite{kumar1996}$^b$      &                 &15.29           & 8.24 \\
Ref.~\cite{bartolomei2010}$^c$ &-0.2251          &15.367          & 8.228 \\
Experimental                   &-0.3$\pm$0.1$^d$ &15.7$\pm$0.3$^e$&8.4$\pm$0.3$^e$  \\
                               &-0.25$^f$        & 15.37$^g$      & 8.22$^g$
\end{tabular}
\end{ruledtabular}
\vskip 0.2cm
$^a$ CBS-CASSCF+1+2\\
$^b$semi-empirical DOSD values\\
$^c$Recommended values: CAS(12e-14o)-ACPF calc. with aug-cc-pV5Z basis set \\
$^d$optical birefringence \cite{Buckingham1968} \\
$^e$vibration rotation Raman spectroscopy \cite{Buldakov1996} \\
$^f$pressure-induced far-infrared spectrum \cite{Cohen1977} \\
$^g$depolarization ratios \cite{Bridge1966}
\end{table}

We have calculated the permanent quadrupole moment $Q$ of O($^3P$) (see Eq.~(\ref{eq:O-quadrup})), with the CASSCF method, in a full-valence active space including 6 electrons and 5 orbitals and an "aug-cc-pVQZ" basis set. Our value $Q= -0.95$~a.u., is in very good agreement with the value obtained with the CCSD(T) using quadruple-$\zeta$ (or higher) basis set \cite{Gutsev1998} of similar quality to the one presently used (Table \ref{tab1}). This indicates that the contribution of dynamical electron correlation effects which are missing in our CASSCF treatment can safely be neglected, provided that sufficiently large basis sets and active space are used. Only static dipole polarizabilities ($\omega=0$) have been reported up to now in the literature, and Table \ref{tab1} shows that our value obtained from Eq.~(\ref{eq:O-pola-pade}) is in satisfactory agreement with other published values.

The available values of the permanent quadrupole moment and static dipole polarizabilities of O$_2$(X$^3\Sigma^-_g$) are shown in Table \ref{tab1bis}. In the present work, the quadrupole moment $q(r)=\left\langle \Lambda_{d}=0\right|\hat{q}_{2}^{0}\left|\Lambda_{d}=0\right\rangle $ is taken from Ref.~\cite{lawson1997}, where it is calculated at the equilibrium distance $r_{e}=2.282$ a.u. and for the lowest vibrational level $v_d=0$ using an harmonic-oscillator approximation. The dynamic dipole polarizabilities are taken from the semi-empirical dipole-oscillator-strength distribution (DOSD) values of Ref.~\cite{kumar1996}. All quantities regarding O$_2$ are in good agreement with the best recommended {\it ab initio} values of Ref.~\cite{bartolomei2010} as well as with experimental values.

\begin{table}
\caption{Long-range non-zero $^{M_{L_a}M'_{L_a}}V^\mathrm{elec}_{52m0}$ electrostatic coefficients
(in atomic units, see Eq.~(\ref{eq:Velec})) for the O($^3P$) + O$_2$(X$^3\Sigma^-_g$) interaction at $r=r_e=2.282$ a.u. and for the lowest vibrational level $v_d=0$ using the data of Table \ref{tab1bis}.}
\label{tab3}
\begin{ruledtabular}
\begin{tabular}{cccrr}
$M_{L_a}$ & $M'_{L_a}$ &  $m$ & $V_5^{\rm elec}(r_e)$ & $V_5^{\rm elec}(v_d=0)$  \\
\hline
 $\pm$1 & $\pm$1 &           0 & -0.721&  -0.647   \\
 $\pm$1 &           0 & $\pm$1 & 0.832&   0.748    \\
 $\pm$1 & $\mp$1 & $\pm$2 & -0.294&  -0.264    \\
           0 & $\pm$1 & $\mp$1 & -0.832&  -0.748    \\
           0 &           0 &           0 & 1.442 &   1.295
\end{tabular}
\end{ruledtabular}
\end{table}

\begin{table}
\caption{Long-range non zero $^{M_{L_a}M'_{L_a}}V^\mathrm{disp}_{6\mathcal{L}_d\mathcal{M}0}$ dispersion
coefficients (in atomic units, see Eq.~(\ref{eq:Edisp})) for the O($^3P$) + O$_2$(X$^3\Sigma^-_g$) interaction at $r=r_e=2.282$~a.u..}
\label{tab4}
\begin{ruledtabular}
\begin{tabular}{ccccr}
$M_{L_a}$ & $M'_{L_a}$ & $\mathcal{M}$ &  $\mathcal{L}_d$ & $V_6^{\rm disp}$  \\
\hline
$\pm$1& $\pm$1&  0&   0& -30.24 \\
 $\pm$1& $\pm$1&   0&   2& -3.665 \\
 $\pm$1&  0&  $\pm$1&   2& -0.253 \\
 $\pm$1&  $\mp$1&  $\pm$2&   2& 0.179 \\
  0& $\pm$1&   $\mp$1&   2& 0.253 \\
  0&  0&  0&   0& -31.511 \\
  0&  0&  0&    2& -4.322
\end{tabular}
\end{ruledtabular}
\end{table}

Tables~\ref{tab3} and \ref{tab4} present the long-range multipolar coefficients at the first (electrostatic) and second (dispersion) orders of the perturbation theory for the matrix elements of O + O$_2$.  The $^{M_{L_a}M'_{L_a}}V^\mathrm{elec}_{52m0}$ and $^{M_{L_a}M'_{L_a}}V^\mathrm{disp}_{6\mathcal{L}_d\mathcal{M}0}$ coefficients are given for each $(M_{L_a}, M'_{L_a})$ pair and for various values of $m$, $\mathcal{L}_d$ and $\mathcal{M}$ at $r=r_e=2.282$ a.u., and also when the relevant quantities are averaged over the $v_d=0$ level of O$_2$. for the former. As already mentioned previously, the maximum value of the dispersion coefficients are about half of those for the O$_2$+O$_2$ long-range interaction of Ref.\cite{bartolomei2010} which represents a good test of the consistency of the calculations.

\begin{figure}
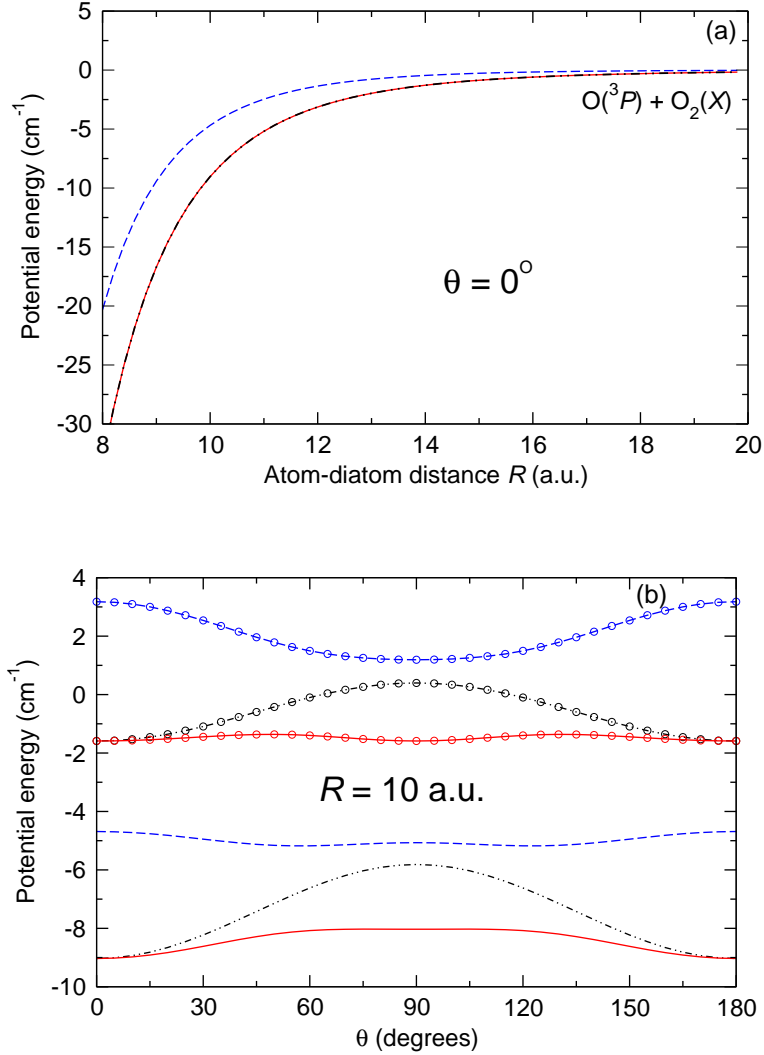

\begin{centering}
\includegraphics[width=100mm]{Fig1a.eps}
\vskip 10mm
\includegraphics[width=100mm]{Fig1b.eps}
\end{centering}
\caption{Asymptotic PESs for the three electronic states $1A'$ (solid line), $1A"$ (dashed-dotted line) and $2A'$ (dashed line) correlated to O($^3P$) + O$_2$(X$^3\Sigma^-_g)$ for $r=r_e=2.282$~a.u.. (a) at $\theta$=0$^\circ$; (b)  at $R=10$~a.u.. In (b), the curves with filled circles refer to the PES when only quadrupole-quadrupole terms are included.}
\label{fig1}
\end{figure}

Asymptotic PESs are obtained after diagonalizing the total interaction potential matrix with elements
\begin{equation}
E_{\mathrm{tot}}^{\, J_aM_aJ'_aM'_a}(r,R,\theta) = \delta_{J_aJ'_a}\delta_{M_aM'_a} E_{0} + E_{\mathrm{elec}}^{\, J_aM_aJ'_aM'_a}(r,R,\theta) + E_{\mathrm{disp}}^{\, J_aM_aJ'_aM'_a}(r,R,\theta),
\label{eq:etot}
\end{equation}
in the subspace spanned by $J_a$ and $M_a$. Figure \ref{fig1} displays one-dimensional cuts of these PESs at $r=r_{e}=2.282$~a.u. either for a fixed bending angle ($\theta=0$, Fig.\ref{fig1}(a)) or at a given O-O$_2$ distance ($R=10$~a.u., Fig.\ref{fig1}(b)). In Fig.~\ref{fig1}(b) we note that the first two states degenerate into a $\Pi$ state for collinear arrangements ($\theta$=0 and 180$^\circ$). Dispersion contributions are noticeable in asymptotic ozone and change slightly the anisotropy of the ground-state potential which, however, remains almost isotropic. The second $A'$ state becomes attractive after inclusion of dispersion energies.

\begin{figure}
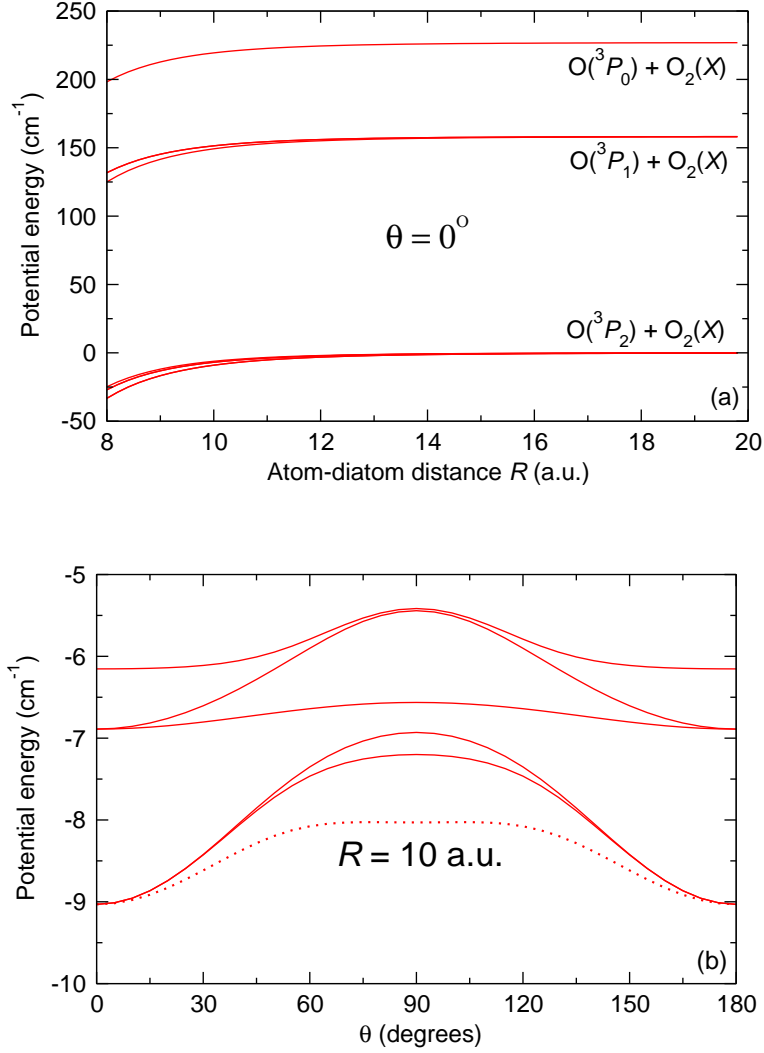

\begin{centering}
\includegraphics[width=100mm]{Fig2a.eps}
\vskip 10mm
\includegraphics[width=100mm]{Fig2b.eps}
\end{centering}
\caption{(a) Asymptotic PESs for the nine spin-orbit states O($^3$P$_{0,1,2}$) + O$_2(X^3\Sigma^-_g)$ at $r=r_e=2.282$~a.u. and $\theta$=0$^\circ$. (b) Asymptotic PESs for the five lowest spin-orbit states correlated to O($^3$P$_2$) + O$_(2X^3\Sigma^-_g)$ at $r=r_e=2.282$~a.u. and $R=10$~a.u.. In (b) the dotted line refer to the ground state PES without fine-structure.}
\label{fig2}
\end{figure}

Inclusion of spin-orbit splitting lifts the degeneracy of the states and twenty-seven states arise from the interaction of O($^3P_{J_a}$) with O$_2$(X$^3\Sigma^-_g$), which reduce to nine states if we neglect the O$_2$ fine-structure (Fig.~\ref{fig2}(a)). The angular dependence of the five states correlated to the O($^3P_{2}$)+O$_2$(X$^3\Sigma^-_g$) limit are displayed for $R=10$~a.u. on Fig.~\ref{fig2}(b), where the influence of the spin-orbit interaction of O($^3P_{2}$) on the anisotropy of the PES is clearly visible.

\section{\label{sec:O2-rovib}Asymptotic potential energy curves between O and O$_{2}$ in a given rovibrational level}

\subsection{Model}

The calculations presented in this section are based on our previous work on
 Cs($6p\,{}^{2}P$)--Cs$_{2}(^{1}\Sigma_{g}^{+},v_{d}=0,N_{d})$ \cite{lepers2010,lepers2011a,lepers2011b,lepers2011c} performed in the context of ultracold gases. The main differences are that the present work deals (i) with a ground-state atom, and (ii) with a triplet molecule ($S_{d}=1$).

When an oxygen atom approaches a rovibrating  O$_{2}$ molecule from large distances, the rotational levels of O$_2$ are coupled by the electric field induced by O. We focus on the derivation of one-dimensional long-range potential energy curves (PECs) for the O--O$_2$ complex depending on $R$, thus  describing the interaction in the frame $xyz$ linked to the complex. In other words we leave out here the mutual rotation of O and O$_2$.

The quantities regarding the oxygen atom are expressed here in the fine-structure basis $|J_aM_a\rangle$ and are identical to those of Section~\ref{sec:FixedGeom}. To stress the differences with the fixed-geometry problem for the diatom, all the quantities
$X$ that are averaged over O$_2$ vibrational wave functions are overlined ($\overline{X}$). Note that the effects of centrifugal distortion on those
wave functions will be ignored.

\subsubsection{Zeroth-order energies and state vectors}

We consider that the O$_{2}$ molecule lies in its ground vibrational level $v_{d}=0$. In addition to the electronic quantum numbers $\Lambda_{d}=0$, $S_{d}=1$ and $\Sigma_{d}$ of O$_{2}$, we introduce the quantum numbers for the (electronic$+$nuclear) orbital angular momentum $N_{d}$, for the orbital$+$spin angular momentum $J_{d}$ ($\left|N_{d}-1\right|\le J_{d}\le N_{d}+1$), and the corresponding projections $M_{d}$, $M_{N_d}$ (and $M_{S_d}$ for $S_d$) on the $z$ axis. Here we limit our study to the $^{16}\textrm{O}_{2}$ isotopologue allowing for odd values of $N_{d}$ only~\cite{vdavoird1987}. All the matrix elements will be given in the fine-structure basis $\left|(N_{d}S_{d})J_{d}M_{d}\right\rangle \equiv\left|N_{d}J_{d}M_{d}\right\rangle $,
connected to the $\left|N_{d}M_{N_d}S_{d}M_{S_d}\right\rangle$ basis by
\begin{equation}
\left|N_{d}J_{d}M_{d}\right\rangle =\sum_{M_{N_d}M_{S_d}}
C_{N_{d}M_{N_d}S_{d}M_{S_d}}^{J_{d}M_{d}}
\left|N_{d}M_{N_d}S_{d}M_{S_d}\right\rangle ,
\label{eq:O2-ChgBase}
\end{equation}
where $C_{N_{d}M_{N_d}S_{d}M_{S_d}}^{J_{d}M_{d}}=\langle N_{d}M_{N_d}S_{d}M_{S_d}|J_{d}M_{d}\rangle$ denotes a Clebsch-Gordan
coefficient.

In its ground electronic state, O$_{2}$ belongs to Hund's case b, \textit{i.e.} $N_{d}$
is considered as a good quantum number. The rotational spectrum is dominated by the free-rotator contribution
$\overline{B}_{v_{d}}\mathbf{N}_{d}^{2}$, with $\overline{B}_{v_{d}=0}=\overline{B}_{0}=1.438$~
cm$^{-1}$ \cite{vdavoird1987}. The matrix element of the spin-spin interaction (Eq.~(\ref{eq:Hss})) are expressed in the $|J_{d}M_{d}\rangle$ basis (see Sec.~III of the supplementary material \cite{epaps})
\begin{eqnarray}
\overline{E}_{SS} & = & \langle J_{d}M_{d}|\hat{H}_{SS}|J_{d}M_{d}\rangle \nonumber \\
 & = & \left(-1\right)^{1+J_{d}}\frac{2\sqrt{30}}{3}\left(2N_{d}+1\right)\overline{\lambda}_{0}\nonumber \\
 &  & \times\left\{ \begin{array}{ccc}
N_{d} & 1 & J_{d}\\
1 & N_{d} & 2\end{array}\right\} \left(\begin{array}{ccc}
N_{d} & 2 & N_{d}\\
0 & 0 & 0\end{array}\right)\,,
\label{eq:Hss-2}
\end{eqnarray}
where the spin-spin coupling constant $\overline{\lambda}_{0}=\overline{\lambda}_{v_{d}=0}=1.983$~cm$^{-1}$ is taken from Ref.~\cite{tinkham1955}. The spin-spin interaction results into the splitting of the fine-structure rotational levels $J_{d}$ inside a given manifold
$N_{d}$. The spin-rotation interaction $\hat{H}_{SR}=\overline{\mu}_{v_{d}=0}\hat{\mathbf{N}}_{d}\cdot\hat{\mathbf{S}}_{d}$ is characterized by a coupling constant $\overline{\mu}_{v_{d}=0}=8.43\times10^{-3}$~cm$^{-1}$~\cite{tinkham1955} much smaller than $\overline{\lambda}_{0}$, and will be neglected in what
follows.

The zeroth-order energy $\overline{E}_{0}$ is the sum of
the atomic spin-orbit, the molecular rigid-rotator and the spin-spin interactions
\begin{equation}
\overline{E}_{0}=\frac{A}{2}\left(J_{a}\left(J_{a}+1\right)-6\right)+\overline{B}_{0}\left(N_{d}\left(N_{d}+1\right)-2\right)+\overline{E}_{SS}\,,\end{equation}
corresponding to unperturbed state vectors $|\overline{\Psi}_{0}\rangle=\left|J_{a}M_{a}\right\rangle \left|N_{d}J_{d}M_{d}\right\rangle $. The origin of energies $\overline{E}_{0}=0$ is fixed to the O$(^3P_2)$+O$_2(v_d=0,N_{d}=1)$ dissociation limit.

\subsubsection{First-order quadrupole-quadrupole interaction}

The quadrupole moment matrix elements of the vibrating and rotating O$_{2}$ molecule is obtained by starting from Eq.~(12) of Ref.~\cite{lepers2011b} written in the $|N_{d}M_{N_d}S_{d}M_{S_d}\rangle$ basis, and by applying the transformation to the fine-structure basis $|N_{d}J_{d}M_{d}\rangle$ (Eq.(\ref{eq:O2-ChgBase}))
\begin{eqnarray}
\langle v_d=0,N_{d}J_{d}M_{d}|\hat{Q}_{2}^{-m}|v_d=0,N'_{d}J'_{d}M'_{d}\rangle  & =
 & \left(-1\right)^{1+J_{d}+J_{d}'-M_{d}'}\overline{q}_{0}\sqrt{\left(2J_{d}+1\right)\left(2J_{d}'+1\right)}\nonumber \\
 &  & \times\sqrt{\left(2N_{d}+1\right)\left(2N_{d}'+1\right)}\left(\begin{array}{ccc}
N_{d} & 2 & N_{d}'\\
0 & 0 & 0\end{array}\right)\nonumber \\
 &  & \times\left\{ \begin{array}{ccc}
N'_{d} & 1 & J'_{d}\\
J_{d} & 2 & N_{d}\end{array}\right\} \left(\begin{array}{ccc}
J_{d} & 2 & J'_{d}\\
-M_{d} & -m & M'_{d}\end{array}\right)\,,
\label{eq:O2-quadrup}
\end{eqnarray}
where we use the expansion of the product of three Clebsh-Gordan coefficients (see Eq.~(2) of the supplementary material \cite{epaps}), and where
\begin{equation}
\overline{q}_{0} \equiv \overline{q}_{v_d=0} = \int_{0}^{+\infty}dr \left(\psi_{v_{d}=0}(r)\right)^{2} q(r)\,
\end{equation}
is the quadrupole operator averaged over the vibrational wave function $\psi_{v_{d}=0}(r)$. Its value $\overline{q}_{0}=-0.2273$~a.u.~is taken
from Ref.~\cite{lawson1997}. The properties of the 3-j and 6-j symbols impose that $N_{d}'=N_{d},N_{d}\pm2$ and $J_{d}'=J_{d},J_{d}\pm1,J_{d}\pm2$.

The $R$-dependent quadrupole-quadrupole matrix element reads
\begin{eqnarray}
 & & \overline{E}_{\mathrm{elec}}^{\, J_a M_a J'_a M'_a N_d J_d M_d N'_d J'_d M'_d}(v_d=0,R) \nonumber\\
 & = & \frac{1}{R^{5}}\sum_{m=-2}^{2}g_{m}(2,2)
 \langle J_{a}M_{a}|\hat{Q}_{2}^{m}|J'_{a}M'_{a}\rangle \nonumber \\
 & & \times \langle v_d=0,N_{d}J_{d}M_{d}|\hat{Q}_{2}^{-m}|v_d=0,N'_{d}J'_{d}M'_{d}\rangle\,,
\label{eq:Eelec-2}
\end{eqnarray}
where $g_{m}(2,2)$ is given by Eq.~(\ref{eq:g-m}).
Note that the atomic quadrupole moment, expressed in the fine-structure basis,
is related to the one in the $|M_{L_a}M_{S_a}\rangle$ basis (see Eq.~(\ref{eq:O-quadrup})),
\begin{eqnarray}
\langle J_{a}M_{a}|\hat{Q}_{2}^{m}|J'_{a}M'_{a}\rangle & = &
\sum_{M_{L_a}M_{S_a}} \sum_{M'_{L_a}M'_{S_a}} C_{1M_{L_a}1M_{S_a}}^{J_a M_a} C_{1M'_{L_a}1M'_{S_a}}^{J'_a M'_a} \nonumber \\
 & \times & \delta_{M_{S_a}M'_{S_a}} \langle 1M_{L_a}|\hat{Q}_{2}^{m}|1M'_{L_a}\rangle \,.
\end{eqnarray}

\subsubsection{Second-order dipole-dipole interaction}

In order to calculate the dispersion term in $R^{-6}$, we use the
approach of Refs.~\cite{lepers2011a,lepers2011b}, adapted with the notations
of the present paper. In the atomic and molecular fine-structure bases, the matrix elements associated with the second-order
dipole-dipole interaction reads
\begin{eqnarray}
 &  & \overline{E}_{\mathrm{disp}}^{\, J_a M_a J'_a M'_a N_d J_d M_d N'_d J'_d M'_d}(R) \nonumber \\
 & = & -\frac{1}{2\pi R^{6}}\sum_{m,m'=-1}^{1}g_{m}(1,1)g_{m'}(1,1)\nonumber \\
 & \times & \int_{0}^{+\infty}d\omega\,{}^{J_a M_a J'_a M'_a}\alpha_{1m1m'}(i\omega)  \nonumber \\
 & & \times \, {}^{N_d J_d M_d N'_d J'_d M'_d}\overline{\alpha}_{1-m1-m'}(i\omega),
\label{eq:Edisp-2}
\end{eqnarray}
where $^{J_a M_a J'_a M'_a}\alpha_{1m1m'}(i\omega)$ and $^{N_d J_d M_d N'_d J'_d M'_d}\overline{\alpha}_{1-m1-m'}(i\omega)$
are the uncoupled dipole polarizabilities at imaginary frequencies,
for the atom and the molecule respectively. Note that these polarizabilities
are related to the ones defined in Ref.~\cite{lepers2011a} according to
$^{xx'}\alpha_{1m1m'}(i\omega)=[(-1)^{m}]{}^{xx'}\alpha_{m,-m'}^{M_{x}M_{x}'}(i\omega)$,
where $x$ stands for the set of atomic or molecular quantum numbers (see Ref.~\cite{lepers2011a}, Eqs.~(13) and (14)).
The polarizabilities of O expressed in the fine-structure basis reads
\begin{eqnarray}
{}^{J_a M_a J'_a M'_a}\alpha_{1m1m'}(i\omega) & = &
\sum_{M_{L_a}M_{S_a}} \sum_{M'_{L_a}M'_{S_a}} C_{1M_{L_a}1M_{S_a}}^{J_a M_a} C_{1M'_{L_a}1M'_{S_a}}^{J'_a M'_a} \nonumber \\
 & \times & \delta_{M_{S_a}M'_{S_a}}\, {}^{M_{L_a}M'_{L_a}}\alpha_{1m1m'}(i\omega) \,.
\end{eqnarray}
where ${}^{M_{L_a}M'_{L_a}}\alpha_{1m1m'}(i\omega)$ are the polarizabilities in the $|M_{L_a}M_{S_a}\rangle$ basis (see Eqs.~(\ref{eq:O-pola-coupl-0}) and (\ref{eq:O-pola-coupl-2})).

For O$_{2}$, the polarizabilities $^{J_d M_d J'_d M'_d}\overline{\alpha}_{1m1m'}$ are calculated in Sec. IV of the supplementary material \cite{epaps}, starting from our previous work \cite{lepers2011a, lepers2011b}. We obtain finally
\begin{eqnarray}
^{N_d J_d M_d N'_d J'_d M'_d}\overline{\alpha}_{1m1m'}(i\omega) & = & \left(-1\right)^{m}\delta_{J_{d}J'_{d}}\delta_{M_{d}M'_{d}}\delta_{m,-m'}\frac{\overline{\alpha}_{0}^{\parallel}(i\omega)+2\overline{\alpha}_{0}^{\bot}(i\omega)}{3}\nonumber \\
 & + & \sqrt{\frac{10}{3}}\sqrt{\left(2N_{d}+1\right)\left(2N'_{d}+1\right)\left(2J_{d}+1\right)\left(2J'_{d}+1\right)}\nonumber \\
 &  & \times\left(-1\right)^{1+J_{d}+J_{d}'-M_{d}'}\left\{ \begin{array}{ccc}
N'_{d} & 1 & J'_{d}\\
J_{d} & 2 & N_{d}\end{array}\right\} \left(\begin{array}{ccc}
N_{d} & 2 & N'_{d}\\
0 & 0 & 0\end{array}\right)\nonumber \\
 &  & \times\left(\overline{\alpha}_{0}^{\parallel}(i\omega)-\overline{\alpha}_{0}^{\bot}(i\omega)\right)\nonumber \\
 &  & \times \left(\begin{array}{ccc}
2 & 1 & 1\\
M'_{d}-M_{d} & m & m'\end{array}\right)\left(\begin{array}{ccc}
J_{d} & 2 & J'_{d}\\
-M_{d} & M_{d}-M'_{d} & M'_{d}\end{array}\right),
\label{eq:O2-alpha-decpl-2}
\end{eqnarray}
where $\overline{\alpha}_{0}^{\parallel}=\overline{\alpha}_{v_{d}=0}^{\parallel}$
and $\overline{\alpha}_{0}^{\bot}=\overline{\alpha}_{v_{d}=0}^{\bot}$
are the parallel and perpendicular polarizabilities of O$_{2}$ in
its vibrational ground level. Equation~(\ref{eq:O2-alpha-decpl-2})
is a sum of two contributions: The first term, proportional
to the so-called isotropic polarizability $(\overline{\alpha}_{0}^{\parallel}+2\overline{\alpha}_{0}^{\bot})/3$,
is diagonal; the second term, which is proportional to the so-called
anisotropic polarizability $\overline{\alpha}_{0}^{\parallel}-\overline{\alpha}_{0}^{\bot}$,
couples the different angular-momentum projections $M_{d}$. The properties of the 3-j and 6-j symbols impose that $N_{d}'=N_{d},N_{d}\pm2$
and $J_{d}'=J_{d},J_{d}\pm1,J_{d}\pm2$.

\subsection{Asymptotic potential energy curves}

\begin{figure}
\centering
\includegraphics[width=8cm]{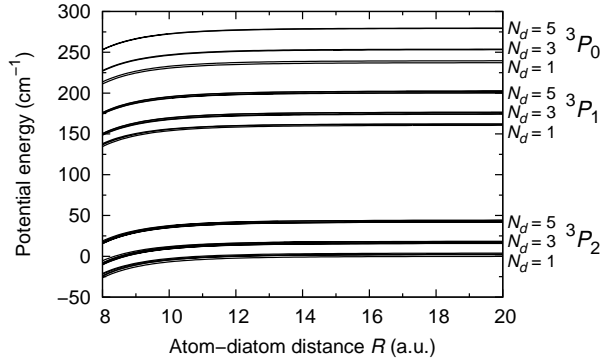}
\caption{\label{fig4} Long-range potential energy curves for the interaction between $^{16}$O$(^3P_{2,1,0})$ and $^{16}$O$_2$($X^3\Sigma_g^-,v_d=0,N_d=1,3,5$), as functions of their mutual separation $R$, for $M=0$ and the $(+)$ reflection symmetry.}
\end{figure}

\begin{figure}
\centering
\includegraphics[width=80mm]{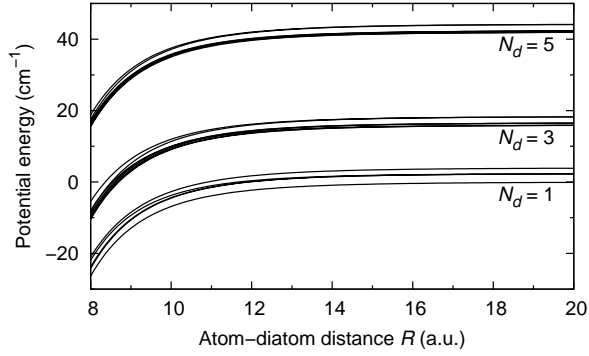}
\caption{\label{fig5} Same as Fig.~\ref{fig4}, but focused on the $^3P_2$ fine-structure level of $^{16}$O.}
\end{figure}

\begin{figure}
\centering
\includegraphics[width=80mm]{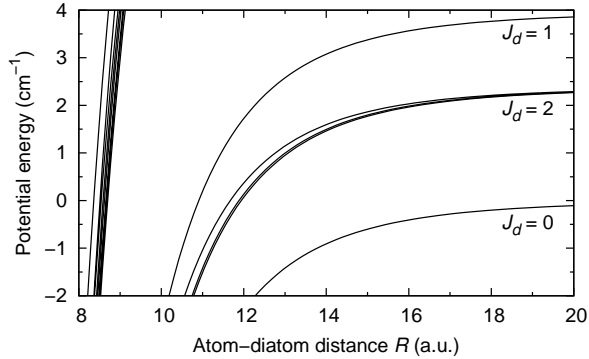}
\caption{\label{fig6} Same as Fig.~\ref{fig4}, but focused on the $^3P_2$ fine-structure level of O, and the $N_d=1$ rotational level of $^{16}$O$_2$.}
\end{figure}

\begin{figure}
\centering
\includegraphics[width=80mm]{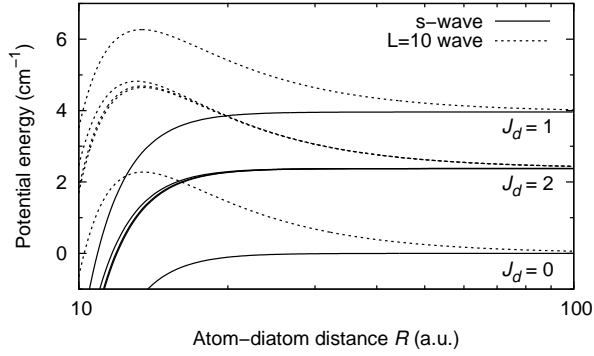}
\caption{\label{fig7} The long-range potential curves of Fig.~\ref{fig6} (solid lines) dressed with a centrifugal term corresponding to a collisional partial wave $L=10$  for the $^{16}$O$_2$-$^{16}$O system. Note that the $R$ axis is logarithmic.}
\end{figure}

The asymptotic PECs are obtained after diagonalizing the potential energy matrix with elements (where we omitted the label $v_d=0$ for simplicity)
\begin{eqnarray}
\overline{E}_{\mathrm{tot}}^{\, J_a M_a J'_a M'_a N_d J_d M_d N'_d J'_d M'_d}(R) & = & \delta_{J_a M_a}\delta_{J'_a M'_a}\delta_{J_d M_d}\delta_{J'_d M'_d} \overline{E}_{0} \nonumber \\
 & + & \overline{E}_{\mathrm{elec}}^{\, J_a M_a J'_a M'_a N_d J_d M_d N'_d J'_d M'_d}(R) \nonumber \\
 & + & \overline{E}_{\mathrm{disp}}^{\, J_a M_a J'_a M'_a N_d J_d M_d N'_d J'_d M'_d}(R) \,,
\label{eq:Etot-2}
\end{eqnarray}
for different values of $R$, and within the subspace determined by each value of the total angular momentum $M=M_a+M_d$ on the $z$ axis \cite{dubernet1994,lepers2010,lepers2011a,lepers2011b,lepers2011c}. For $M=0$, the eigenvectors of (\ref{eq:Etot-2}) are also characterized by a given reflection symmetry through any plane containing the $z$ axis. Following the expression of the reflection operator $\sigma_{xz}$ for two atoms (see for instance Eq.~(3.11) of Ref.~\cite{chang1967}), we assign the $\pm$ symmetry to eigenvectors corresponding to the linear combinations $|M_{a},-M_{a}\rangle \pm(-1)^{L_a+J_a+N_d+J_d} |-M_{a}M_{a}\rangle$. On Figs.~\ref{fig4}--\ref{fig7}, we display long-range potential curves belonging to the $M=0^+$ symmetry. The curves belonging to other symmetries have the same appearance, except that some asymptotic channels may not be allowed for a given value of $M$, \textit{e.g.} $J_a=J_d=0$ for $M=1$. All the curves are strongly attractive, due to the dominance of the attractive van der Waals term in the range of O--O$_2$ distances that we consider here (see Tables \ref{tab3} and \ref{tab4} for an illustration of this point).

As the spin-orbit splitting of O is much larger than the rotational splitting of O$_2$ (at least for the $N_d$ lowest levels), we see in Fig.~\ref{fig4} that the corresponding  fine-structure PEC manifolds are almost decoupled from each other. This situation corresponds to Hund's-like case "1C" defined in \cite{dubernet1994}, also valid in the Cs*+Cs$_2$ system \cite{lepers2011c}. For the sake of clarity, we have restricted our plot to the first three (odd-parity) rotational levels of $^{16}$O$_2$ ($N_d=1,3,5$), even if in principle all the rotational levels up to infinity should be included in the calculation (see a discussion about this point in \cite{lepers2011b}). However, even at the low-$R$ limit of our calculations, the different $N_d$ levels remain almost uncoupled. As an illustration, at $R=8$~a.u., the PECs of Figs.~\ref{fig4}--\ref{fig6} conserve at least $95\%$ of their asymptotic $N_d$ character. In other words, in the range of O--O$_2$ distances that we consider, the oxygen molecule rotates almost freely.

%Figures \ref{fig5} and \ref{fig6} are zooms on Figure \ref{fig4} on the $(J_a=2)$ and $(J_a=2, N_d=1)$ manifolds respectively. We can see that the range of energy that we describe here goes down to 30 cm$^{-1}$ below the lowest dissociation limit. Therefore with our curves, we can expect to describe highly excited vibrational levels along $R$, mostly located in a range of O--O$_2$ distances larger than 8 Bohr radii.

The PECs obtained for other O$_2$ isotopologues, not shown here, have similar features as Figs.~\ref{fig4}--\ref{fig6}. But caution has to be taken on which $N_d$ values are allowed. For $^{18}\textrm{O}_2$, the allowed $N_d$ values are the odd ones, whereas for $^{17}\textrm{O}_2$, the allowed values are the even/odd ones, if the total nuclear spin is odd/even. For mixed isotopologues, that is $^{16}\textrm{O}^{17}\textrm{O}$, $^{16}\textrm{O}^{18}\textrm{O}$ and $^{17}\textrm{O}^{18}\textrm{O}$, all values of $N_d$ are possible. But since the Hamiltonian of Eq.~(\ref{eq:Etot-2}) conserves the parity of $N_d$, even and odd values can still be separated from each other.

In order to get an insight into the collisional dynamics between O and O$_2$, we have added to our curves a centrifugal term $L(L+1)/2\mu R^2$, where $L$ is the partial wave of the collision that we took equal to 10, and $\mu$ the reduced mass of the O--O$_2$ system (Fig.~\ref{fig7}). This term competes with the attractive van der Waals interaction, and creates a centrifugal barrier of about 2~cm$^{-1}$, or about 3~Kelvin. This temperature being much lower than even the lowest temperatures of ozone samples, we can conclude that many partial waves should be included in a full quantum treatment of the O--O$_2$ collision.

\section{\label{sec:Discussion}Discussion: how do the two approaches connect to each other?}

In the previous sections, we have calculated the matrix elements of the potential energy operator of the long-range interaction between ground state O and O$_2$, yielding its eigenvalues after diagonalization in the appropriate configuration subspace: either for fixed geometries of the O--O$_{2}$ complex leading to ($R$,$\theta$,$r\equiv r_e$)-dependent electronic PESs (Section~\ref{sec:FixedGeom}), or for a given rovibrational state of O$_2$ leading to $R$-dependent PECs (Section~\ref{sec:O2-rovib}). It is clear that those two adiabatic potential energy sets are not directly related to each other; instead, the matrix elements of the latter approach are obtained by averaging those of the former approach on the $r$ and $\theta$ coordinates of O$_2$, as briefly addressed in Ref.~\cite{dubernet1994}. This is easily demonstrated on the matrix elements of the quadrupole-quadrupole interaction. Equation (\ref{eq:O2-quadrup}) shows that the ro-vibrationally averaged matrix element of the diatom quadrupole moment reads
\begin{equation}
\left\langle
v_{d}N_{d}J_{d}M_{d}\left|\hat{Q}_{2}^{-m}\right|v_{d}N_{d}'J_{d}'M_{d}
'\right\rangle
 = \overline{q}_{v_d} \int_{0}^{\pi}d\theta\sin\theta\psi_{N_{d}J_{d}M_{d}}(\theta)d_{-m0}^{2}(\theta)\psi_{N_{d}'J_{d}'M_{d}'}(\theta) \,,
\label{eq:disc-quadrup}
\end{equation}
where
\begin{equation}
\psi_{N_{d}J_{d}M_{d}}(\theta)=\sqrt{\frac{2N_{d}+1}{2}}\sum_{M_{N_d}M_{Ns}}\langle N_{d}M_{N_d}S_{d}M_{S_d}|J_{d}M_{d}\rangle d_{M_{N_d}0}^{N_{d}}(\theta)
\end{equation}
is the angular part of the wave function of the fine-structure rotational level. We thus have
\begin{eqnarray}
\overline{E}_{\mathrm{elec}}^{\, J_a M_a J'_a M'_a N_d J_d M_d N'_d J'_d M'_d}(R)  =  \int_{0}^{+\infty}dr\left(\psi_{v_{d}}(r)\right)^{2} \int_{0}^{\pi}d\theta\sin\theta\psi_{N_{d}J_{d}M_{d}}(\theta)\psi_{N_{d}'J_{d}'M_{d}'}(\theta) E_{\mathrm{elec}}^{\, J_aM_aJ'_aM'_a}(R,r,\theta)\,.
\label{eq:Moy-Eelec}
\end{eqnarray}

The situation for the second-order dipole-dipole interaction is slightly more subtle. In Sec.~V of the supplementary material \cite{epaps}, we demonstrate that as far as the rotation of O$_{2}$ is concerned, the passage from one approach to the other is made by averaging $E_{\mathrm{disp}}$ over the rotational wave function $\psi_{N_{d}J_{d}M_{d}}(\theta)$, which can be schematically summarized as follows
\begin{equation}
d_{-\mathcal{M}0}^{\mathcal{L}_{d}}(\theta)\mathrm{\, [in\,}E_{\mathrm{disp}}\mathrm{]}
\to \int_{0}^{\pi}d\theta\sin\theta\psi_{N_{d}J_{d}M_{d}}(\theta)d_{-\mathcal{M}0}^{\mathcal{L}_{d}}(\theta)\psi_{N_{d}'J_{d}'M_{d}'}(\theta) \mathrm{\,[in\,}\overline{E}_{\mathrm{disp}}\mathrm{]}\,,
\label{eq:Ang-Average}
\end{equation}
where $E_{\mathrm{disp}} = E_{\mathrm{disp}}^{J_a M_a J'_a M'_a}$ and $\overline{E}_{\mathrm{disp}} = \overline{E}_{\mathrm{disp}}^{\, J_a M_a J'_a M'_a N_d J_d M_d N'_d J'_d M'_d}$ are given by Eqs.~(\ref{eq:Edisp}) and (\ref{eq:Edisp-2}), respectively.

The drawback which can occur when taking in account the vibration of O$_{2}$ is illustrated by considering the static parallel electronic polarizability $\alpha^\parallel(\omega=0;r)$
\begin{equation}
\alpha^\parallel(\omega=0;r) = 2\sum_{e\neq X} \frac{\mu_{Xe}^2(r)}{V_e(r)-V_X(r)}\,,
\label{eq:O2-alpha-elec}
\end{equation}
where $V_X(r)$ and $V_e(r)$ are the potential energy curves of the $^3\Sigma_g^-$ ground state and of the excited electronic states, respectively, and $\mu_{Xe}(r)$ the ($r$-dependent) transition dipole moment between states $X$ and $e$. Eq.~(\ref{eq:O2-alpha-elec}) is a strict application of the Born-Oppenheimer approximation, \textit{i.e.} the molecular response to an electric field is only due to the electrons. The most straightforward way to take into account the vibration of O$_2$ would be to average (\ref{eq:O2-alpha-elec}) over the vibrational wave function,
\begin{equation}
\langle\alpha^\parallel(r)\rangle_{v_d} = 2\sum_{e\neq X}
\int_0^{+\infty}dr \psi_{v_d}(r) \frac{\mu_{Xe}^2(r)}{V_e(r)-V_X(r)} \psi_{v_d}(r)\,.
\label{eq:O2-alpha-vib-1}
\end{equation}
However, one sees that Eq.~(\ref{eq:O2-alpha-vib-1}) differs from the polarizability $\overline{\alpha}_{v_d}^{\parallel}$ associated with the vibrational level $v_d$, which reads
\begin{equation}
\overline{\alpha}_{v_d}^{\parallel} = 2\sum_{e,v'_d\neq X,v_d}
\frac{\langle v'_d|\mu_{Xe}(r)|v_d\rangle^2}{E_{ev'_d}-E_{Xv_d}}\,,
\label{eq:O2-alpha-vib-2}
\end{equation}
where $E_{ev'_d}$ and $E_{Xv_d}$ are the electronic and vibrational energy levels of the field-free molecule. Note that $v'_d$ can also stand for a continuum
state, in which case the discrete sum in Eq.~(\ref{eq:O2-alpha-vib-2}) becomes an integral.  As pointed out in Ref.~\cite{bishop1990}, Eqs.(\ref{eq:O2-alpha-vib-1}) and (\ref{eq:O2-alpha-vib-2}) are identical if one neglects the diatom zero-point energy, \textit{i.e.} $E_{Xv_d}=E_X(r=r_e)$ and $E_{ev'_d}=E_e(r=r_e)$, where $r_e$ is the O$_2$ equilibrium distance in its ground state.

\begin{figure}
\centering
\includegraphics[width=80mm]{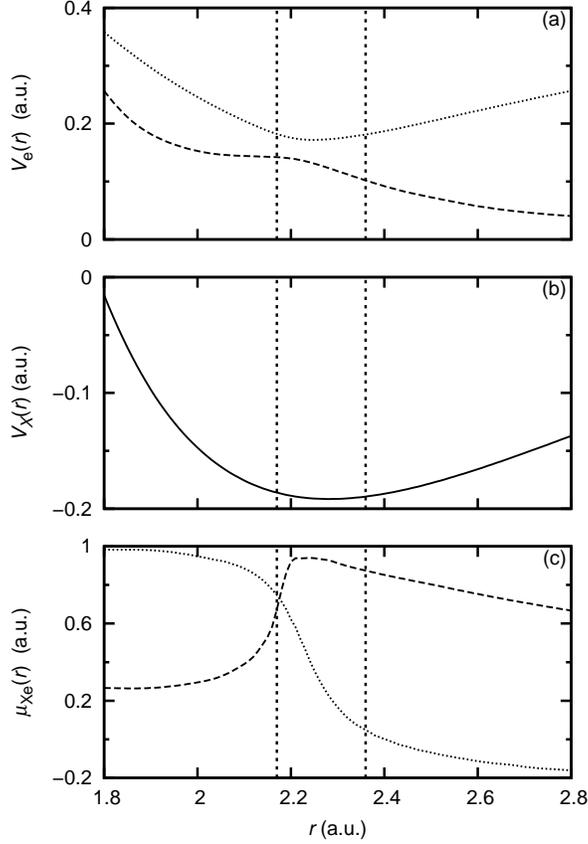}
\caption{\label{fig:O2-PES} Potential energy curves of O$_2$: (a) for the $B^3\Sigma_u^-$ state (dashed lines) and the $E^3\Sigma_u^-$ state (dotted lines), (b) for the $X^3\Sigma_g^-$ state. (c) transition dipole moments: $B\leftarrow X$ (dashed lines), $E\leftarrow X$ (dotted lines). Between the two vertical lines, the wave-function of the $X,v_d=0$ level is more than half its maximum value.}
\end{figure}

To check the validity of this approximation is appropriate for O$_2$, we have considered the two lowest excited $B^3\Sigma_u^-$ and $E^3\Sigma_u^-$ electronic states in the sums above (Fig.~\ref{fig:O2-PES}). The  $X$ and $B$ PECs, as well as the transition dipole moment $\mu_{XB}(r)$ are taken from Refs.~\cite{allison1986,friedman1990}. The PEC for the pure $E$ Rydberg state is taken from Ref.~\cite{wang1987}, and we have extended it for $2.8a_0<r<3a_0$ by shifting down the experimental potential curve of O$_2^+(X^2\Pi_g)$~\cite{singh1966}. The transition dipole moment between $X$ and $E$ has been extracted from Ref.~\cite{wang1987}. As described in Ref.\cite{spelsberg1998}, the $B$ and $E$ states exhibit an avoided crossing around 2.3$a_0$ close to the $X$ equilibrium distance $r_e$, due to a valence-Rydberg exchange of character. The $B$ state is steep and purely repulsive around $r_e$, while the minimum of the $E$ state is located at $r=2.25a_0$, namely close to $r_e$.

The numerical values of Eqs.(\ref{eq:O2-alpha-elec}-\ref{eq:O2-alpha-vib-2}) obtained within this three-state model are presented in Table \ref{tab5}. The electronic polarizability of the $X$ state $\alpha^\parallel(r_e)=$5.75~a.u. amounts for 40\% of the total parallel polarizability \cite{kumar1996}. The three determinations are almost identical, despite the particular shape of the PECs. The partial contribution of each transition, ($B\leftarrow X$) and ($E\leftarrow X$) show a discrepancy between $\alpha^\parallel(r)$ on the one hand, and $\langle\alpha^\parallel(r)\rangle_0$ and $\overline{\alpha}_0^{\parallel}$ on the other hand, which is due to the sudden changes in potential curves and transition dipole moment within the spatial extension of the $X,v_d=0$ level. The fact that these differences compensate each other is due to the exchange in character between the two excited electronic states. In contrast, inside each electronic transition, $\langle\alpha^\parallel(r)\rangle_{v_d}$ and $\overline{\alpha}_{v_d}^{\parallel}$ are always nearly equal. On a physical point of view, it confirms that the response to an external electric field is, to a very large extent, due to the electrons, and not to the nuclei, which justifies the use of the Born-Oppenheimer approximation.

%The electronic polarizability $\alpha^\parallel(r)$ was calculated by applying Eq.~(\ref{eq:O2-alpha-elec}) at $r=r_e$. The result, for our three-state model, is 5.75~a.u., that is 40\% of the total parallel polarizability \cite{kumar1996}. For $\langle\alpha^\parallel(r)\rangle_{v_d}$, we have calculated the wave function of the $v_d=0$ level with our Discrete Variable Representation (DVR) code. For $\overline{\alpha}_{v_d}^{\parallel}$, the wave functions of $X$, $B$ and $E$ states have been calculated separately with our DVR code. Note that the continuum wave functions of state $B$ have been treated as a discretized continuum. All the integrals involving the transition dipole moments have been calculated in a second step.

\begin{table}
\caption{Static parallel polarizability of O$_2$ calculated with three electronic states (see text): $\alpha^\parallel(r_e)$ is purely electronic polarizability, evaluated at the O$_2$ equilibrium distance (see Eq.~(\ref{eq:O2-alpha-elec})),
$\langle\alpha^\parallel(r)\rangle_{v_d=0}$ is the $r$-dependent electronic polarizability averaged over the ground-level vibrational wave function (see Eq.~(\ref{eq:O2-alpha-vib-1})), and
$\overline{\alpha}_{v_d=0}^{\parallel}$ is purely vibrational (see Eq.~(\ref{eq:O2-alpha-vib-2})). We present the contribution of individual electronic transitions, $B\leftarrow X$ and $E\leftarrow X$, to the polarizability.
\label{tab5}}
\begin{ruledtabular}
\begin{tabular}{cccc}
 Transition & $\alpha^\parallel(r_e)$ &
 $\langle\alpha^\parallel(r)\rangle_{v_d=0}$ &
 $\overline{\alpha}_{v_d=0}^{\parallel}$ \\
\hline
 $B\leftarrow X$ & 5.47 & 5.16 & 5.15 \\
 $E\leftarrow X$ & 0.29 & 0.63 & 0.63 \\
 Total            & 5.75 & 5.79 & 5.78 \\
\end{tabular}
\end{ruledtabular}
\end{table}

Our result tends to prove that the vibrational part of the energy (see Eq.~(\ref{eq:O2-alpha-vib-2})) can be neglected, whether the excited state PEC has a potential well or is purely repulsive. In the first situation, which corresponds to the $E\leftarrow X$ transition, the ground vibrational level $v_d=0$ has significant Franck-Condon factor (higher than $10^{-4}$) with the $E$ vibrational levels ranging from $v=0$ to 4. The vibrational energies (with respect to the corresponding minimum electronic energies) associated with $X,v_d=0$ and $E,v=4$ are 791~cm$^{-1}$ and 10300~cm$^{-1}$ respectively. So they are very small compared to the difference in electronic energy, which equals to 80000~cm$^{-1}$ at $r=r_e$.
The ground vibrational level $v_d=0$ significantly overlaps with continuum states of the purely repulsive $B$ state whose classical turning point is located in the Franck-Condon region. This turning point region is precisely where the vibrational, \textit{i.e.} kinetic part of the energy is small.
In the present case, we only have the O$_2$ polarizabilities at the equilibrium distance $r_e$. In consequence, we assumed that the relationship between the fixed-geometries and the ro-vibrationally averaged approach, given for $E_\mathrm{elec}$ in Eq.~(\ref{eq:Moy-Eelec}), can be extended to $E_\mathrm{disp}$, and hence for $E_\mathrm{tot}$, within a good approximation.

\section{\label{sec:Conclusion}Conclusion}

In this article, we have characterized the potential energy surfaces (PESs) of the ozone molecule at large distances, where O and O$_2$ weakly interact with each other, using two different approaches. Our calculations are based on the multipolar expansion of the electrostatic potential energy, which is expressed in terms of the electric properties of O and O$_2$.

Firstly, we have calculated the 27 asymptotic electronic PESs correlated to the O$(^3P)$--O$_2(X^3\Sigma_g^-)$ dissociation limit including spin-orbit couplings. All the PESs are found attractive due to the dominant isotropic van der Waals interaction. For all spin-orbit states, the minimum energy is observed for a linear configuration of the three oxygen atoms. These long-range PESs can readily be connected at short distances to the most recent ones obtained by high-level {\it ab initio} calculations \cite{dawes2011}, in order to derive global PESs for all values of the chosen internal coordinates. The overlap of the long-range and {\it ab initio} PESs on a large enough range of distances in the region of the LeRoy radius should avoid the tedious calculation of higher order terms in the multipolar expansion.

Secondly, we have considered the interaction between an O$_2(X^3\Sigma_g^-)$ molecule in a given spin-orbit state, vibrating in the ground level and rotating in a low level and an O$(^3P)$ atom.  We obtained one-dimensional potential energy curves (PECs) dominated by the isotropic van der Waals interaction. They are nearly parallel, which indicates that, when O and O$_2$ approach each other, the rotation of O$_2$ is only weakly hindered by the presence of O at distances larger than the LeRoy radius. It is worthwhile to note that this hypothesis is generally assumed in models characterizing the transition state, which is located at smaller distances, in the O+O$_2$ exchange reaction \cite{hathorn2000,gao2001, gao2002}. However, such an hypothesis of weakly hindered rotation is not always appropriate, as demonstrated for instance in the long-range interaction between a cesium atom and a ground state Cs$_2$ molecule \cite{lepers2011c}.

The one-dimensional PEC above can be smoothly matched to an arbitrary potential curve below the LeRoy radius so that vibrational energies and wavefunctions (along the O--O$_2$ axis) can be calculated in the frame attached to the trimer ({\it i.e.}~neglecting the mutual rotation of O and O$_2$. They will be of relevance for estimating the density of vibrational levels of the O--O$_2$ system close to the dissociation limit, namely with binding energy smaller than the remaining potential energy $V(R_{LR}) \approx30$~cm$^{-1}$ around the LeRoy radius. Such wavefunctions will be mostly concentrated at large distances, since the potential energy around the O$_3$ global minimum is by far larger than $V(R_{LR})$. Once experimental energies are available near the dissociation limit, one could tune the potential energy in the inner zone in order to match the calculated energies to the observed one, and thus characterize the related energy levels with quantum numbers of the O--O$_2$ system.

Furthermore, such PECs could be of interest for the study of O--O$_2$ collisions in the cold ($T\lesssim1$ K) or ultracold ($T\lesssim1$ mK) regimes. Several theoretical studies already aimed at modeling collisions involving oxygen \cite{avdeenkov2001, volpi2002, tilford2004, perez-rios2009, perez-rios2011}, while several experimental investigations suggested possible ways to create beams of cold atomic or molecular oxygen \cite{patterson2007,narevicius2008,bucicov2008}. In particular, we can expect such collisions to exhibit {\it rotational} unconventional isotopic effects. If the difference in rotational energy compensates the difference in vibrational zero-point energy, we obtain two quasi-degenerate channels of different isotopologues. In addition, the nuclear-spin symmetry can also play an important part in the rotational isotopic effects, as it forbids some rotational levels.
For example, the channels $^{17}\mathrm{O}+^{16}\mathrm{O}_2(N_d=3)$ and
$^{16}\mathrm{O}+^{16}\mathrm{O}^{17}\mathrm{O}(N_d=4)$
on the one hand, and
$^{18}\mathrm{O}+^{17}\mathrm{O}^{18}\mathrm{O}(N_d=3)$ and $^{17}\mathrm{O}+^{18}\mathrm{O}_2(N_d=4
)$ on the other hand, are only separated by 0.2 cm$^{-1}$; but the second situation is not allowed, as $^{18}\mathrm{O}_2$ has no even rotational levels.

It is worthwhile to mention that in contrast with similar studies devoted to low energy atom exchange reactions in atom-diatom systems (for instance Al+O$_2$ \cite{reignier1998}, C+OH \cite{jorfi2010}, or O+OH \cite{stoecklin2012}), the adiabatic capture centrifugal theory~\cite{levine1987} conducted in the sudden approximation (ACCSA)~\cite{clary1984} relying on the unique knowledge of the long-range asymptotic ground state potential energy of the atom-diatom system cannot be used in the ozone case, due to the presence of the transition state playing a role at distances lower than the LeRoy radius, and thus beyond the validity of the present approach. A calculation within this model would undoubtedly lead to an overestimation of the atom exchange rate.

\section*{Acknowledgments}

We thank Pr.~Vladimir Tyuterev for attracting our attention on the ozone problem. This work was done with the support of ``{}Triangle de la Physique" under contract 2008-007T-QCCM (Quantum Control of Cold Molecules).

%\bibliographystyle{unsrt}
%\bibliography{Biblio}

\begin{thebibliography}{10}

\bibitem{schinke2006}
R.~Schinke, S.~Y. Grebenshchikov, M.~V. Ivanov, and P.~Fleurat-Lessard.
\newblock Dynamical studies of the ozone isotope effect: A status report.
\newblock {\em Annu. Rev. Phys. Chem.}, 57:625--661, 2006.

\bibitem{thiemens1983}
M.H. Thiemens and J.E. Heidenreich.
\newblock The mass-independent fractionation of oxygen: A novel isotope effect
  and its possible cosmochemical implications.
\newblock {\em Science}, 219:1073, 1983.

\bibitem{mauersberger1987}
K.~Mauersberger.
\newblock Ozone isotope measurements in the stratosphere.
\newblock {\em Geophys. Res. Lett.}, 14:80--83, 1987.

\bibitem{anderson1997}
S.M. Anderson, D.~H{\"u}lsebusch, and K.~Mauersberger.
\newblock Surprising rate coefficients for four isotopic variants of
  o$+$o$_2+$m.
\newblock {\em J. Chem. Phys.}, 107:5385--5392, 1997.

\bibitem{janssen2001}
C.~Janssen, J.~Guenther, K.~Mauersberger, and D.~Krankowsky.
\newblock Kinetic origin of the ozone isotope effect: a critical analysis of
  enrichments and rate coefficients.
\newblock {\em Phys. Chem. Chem. Phys.}, 3(21):4718--4721, 2001.

\bibitem{mauersberger2005}
K.~Mauersberger, D.~Krankowsky, C.~Janssen, and R.~Schinke.
\newblock Assessment of the ozone isotope effect.
\newblock {\em Adv. At. Mol. Opt. Phys.}, 50:1--54, 2005.

\bibitem{janssen1999}
C.~Janssen, J.~Guenther, D.~Krankowsky, and K.~Mauersberger.
\newblock Relative formation rates of $^{50}$o$_3$ and $^{52}$o$_3$ in
  $^{16}$o--$^{18}$o mixtures.
\newblock {\em J. Chem. Phys.}, 111:7179--7182, 1999.

\bibitem{gao2001}
Y.Q. Gao and R.A. Marcus.
\newblock Strange and unconventional isotope effects in ozone formation.
\newblock {\em Science}, 293:259--263, 2001.

\bibitem{gao2002}
Y.Q. Gao and R.A. Marcus.
\newblock On the theory of the strange and unconventional isotopic effects in
  ozone formation.
\newblock {\em J. Chem. Phys.}, 116:137--154, 2002.

\bibitem{grebenshchikov2003}
S.Y. Grebenshchikov, R.~Schinke, P.~Fleurat-Lessard, and M.~Joyeux.
\newblock van der {W}aals states in ozone and their influence on the threshold
  spectrum of {O} ($x^1a_1$). i. bound states.
\newblock {\em J. Chem. Phys.}, 119:6512, 2003.

\bibitem{babikov2003}
D.~Babikov.
\newblock Entrance channel localized states in ozone: Possible application to
  helium nanodroplet isolation spectroscopy.
\newblock {\em J. Chem. Phys.}, 119:6554, 2003.

\bibitem{lee2004}
H.S. Lee and J.C. Light.
\newblock Vibrational energy levels of ozone up to dissociation revisited.
\newblock {\em J. Chem. Phys.}, 120:5859, 2004.

\bibitem{babikov2003b}
D.~Babikov, B.K. Kendrick, R.B. Walker, R.T. Pack, P.~Fleurat-Lesard, and
  R.~Schinke.
\newblock Metastable states of ozone calculated on an accurate potential energy
  surface.
\newblock {\em J. Chem. Phys.}, 118:6298--6308, 2003.

\bibitem{grebenshchikov2009}
S.~Y. Grebenshchikov and R.~Schinke.
\newblock Towards quantum mechanical description of the unconventional
  mass-dependent isotope effect in ozone: Resonance recombination in the strong
  collision approximation.
\newblock {\em J. Chem. Phy.}, 131:181103, 2009.

\bibitem{fleurat2003}
P.~Fleurat-Lessard, S.Y. Grebenshchikov, R.~Siebert, R.~Schinke, and
  N.~Halberstadt.
\newblock Theoretical investigation of the temperature dependence of the
  o+o$_2$ exchange reaction.
\newblock {\em J. Chem. Phys.}, 118:610, 2003.

\bibitem{ivanov2012}
M.V. Ivanov and D.~Babikov.
\newblock Efficient quantum-classical method for computing thermal rate
  constant of recombination: Application to ozone formation.
\newblock {\em J. Chem. Phys.}, 136:184304, 2012.

\bibitem{siebert2001}
R.~Siebert, R.~Schinke, and M.~Bittererov{\'a}.
\newblock Spectroscopy of ozone at the dissociation threshold: Quantum
  calculations of bound and resonance states on a new global potential energy
  surface.
\newblock {\em Phys. Chem. Chem. Phys.}, 3:1795--1798, 2001.

\bibitem{siebert2002}
R.~Siebert, P.~Fleurat-Lessard, R.~Schinke, M.~Bittererov{\'a}, and S.C.
  Farantos.
\newblock The vibrational energies of ozone up to the dissociation threshold:
  Dynamics calculations on an accurate potential energy surface.
\newblock {\em J. Chem. Phys.}, 116:9749--9767, 2002.

\bibitem{rosmus2002}
P.~Rosmus, P.~Palmieri, and R.~Schinke.
\newblock The asymptotic region of the potential energy surfaces relevant for
  the o $(p)+$o$_2(x^3\sigma_g^-$)$\rightleftharpoons$o$_3$ reaction.
\newblock {\em J. Chem. Phys.}, 117:4871--4877, 2002.

\bibitem{holka2010}
F.~Holka, P.G. Szalay, T.~M{\"u}ller, and V.G. Tyuterev.
\newblock Toward an improved ground state potential energy surface of ozone.
\newblock {\em J. Phys. Chem. A}, 114:9927--9935, 2010.

\bibitem{dawes2011}
R.~Dawes, P.~Lolur, J.~Ma, and H.~Guo.
\newblock Communication: Highly accurate ozone formation potential and
  implications for kinetics.
\newblock {\em J. Chem. Phys.}, 135:081102, 2011.

\bibitem{tashiro2003}
M.~Tashiro and R.~Schinke.
\newblock The effect of spin--orbit coupling in complex forming o($^3p$)+o$_2$
  collisions.
\newblock {\em J. Chem. Phys.}, 119:10186, 2003.

\bibitem{bussery-honvault2008}
B.~Bussery-Honvault, F.~Dayou, and A.~Zanchet.
\newblock Long-range multipolar potentials of the 18 spin-orbit states arising
  from the {C} ($^3{P}$) + {OH}({X}$^2{\Pi}$) interaction.
\newblock {\em J. Chem. Phys.}, 129:234302, 2008.

\bibitem{bussery-honvault2009}
B.~Bussery-Honvault and F.~Dayou.
\newblock {Si} ($^3{P}$) + {OH} ({X}$^2{\Pi}$) interaction: Long-range
  multipolar potentials of the eighteen spin-orbit states.
\newblock {\em J. Phys. Chem. A}, 113(52):14961, 2009.

\bibitem{lepers2010}
M.~Lepers, O.~Dulieu, and V.~Kokoouline.
\newblock Photoassociation of a cold atom-molecule pair: long-range
  quadrupole-quadrupole interactions.
\newblock {\em Phys. Rev. A}, 82:042711, 2010.

\bibitem{leroy1974}
R.J. LeRoy.
\newblock Long-range potential coefficients from {RKR} turning points: {$C_6$}
  and {$C_8$} for {$B(^{3}{\Pi}_{{Ou}^+})$}-state {Cl}$_2$, {Br}$_2$, and
  {I}$_2$.
\newblock {\em Can. J. Phys.}, 52:246, 1974.

\bibitem{bartolomei2010}
M.~Bartolomei, E.~Carmona-Novillo, J.~Campos-Martinez M.I.~Hernandez, and
  R.~Hernandez-Lamoneda.
\newblock {\em J. Comput. Chem.}, 32:279--290, 2010.

\bibitem{tinkham1955}
M.~Tinkham and M.~W.~P. Strandberg.
\newblock Theory of the fine structure of the molecular oxygen ground state.
\newblock {\em Phys. Rev.}, 97:937--951, 1955.

\bibitem{varshalovich1988}
D.~A. Varshalovich, A.~N. Moskalev, and V.~K. Khersonskii.
\newblock {\em Quantum theory of angular momentum}.
\newblock Leningrad, 1988.

\bibitem{spelsberg1993}
D.~Spelsberg, T.~Lorenz, and W.~Meyer.
\newblock Dynamic multipole polarizabilities and long range interaction
  coefficients for the systems {H, Li, Na, K, He, H$^-$, H$_2$, Li$_2$,
  Na$_2$}, and {K$_2$}.
\newblock {\em J. Chem. Phys.}, 99:7845, 1993.

\bibitem{langhoff1970}
P.W. Langhoff and M.~Karplus.
\newblock Pad{\'e} approximants for two-and three-body dipole dispersion
  interactions.
\newblock {\em J. Chem. Phys.}, 53:233--250, 1970.

\bibitem{epaps}
{\em EPAPS no. NNN}.

\bibitem{stoecklin2012}
T.~Stoecklin, B.~Bussery-Honvault, P.~Honvault, and F.~Dayou.
\newblock Asymptotic potentials and rate constants in the adiabatic capture
  centrifugal sudden approximation for {X + OH$(X^2\Pi)\to$ OX + H$(^2S)$}
  reactions where {X = O$(^3P)$, S$(^3P)$ or N$(^4S)$}.
\newblock {\em Comp. Theor. Chem.}, 990:39--46, 2012.

\bibitem{Gutsev1998}
G.L. Gutsev, P.~Jena, and R.J. Bartlett.
\newblock {\em Chem. Phys. Lett.}, 291:547, 1998.

\bibitem{Das1998}
A.K. Das and A.J. Thakkar.
\newblock {\em J. Phys. B}, 31:2215, 1998.

\bibitem{Medved2000}
M.~Medved, P.W. Fowler, and J.M. Hutson.
\newblock {\em Mol. Phys.}, 98:453, 2000.

\bibitem{lawson1997}
D.~B. Lawson and J.~F. Harrison.
\newblock Distance dependence and spatial distribution of the molecular
  quadrupole moments of {H}$_2$, {N}$_2$, {O}$_2$, and {F}$_2$.
\newblock {\em J. Phys. Chem. A}, 101(26):4781--4792, 1997.

\bibitem{kumar1996}
A.~Kumar, W.~J. Meath, P.~B{\"u}ndgen, and A.~J. Thakkar.
\newblock Reliable anisotropic dipole properties, and dispersion energy
  coefficients, for o evaluated using constrained dipole oscillator strength
  techniques.
\newblock {\em J. Chem. Phys.}, 105:4927--4937, 1996.

\bibitem{Buckingham1968}
D.A.~Dummur A.D.~Buckingham, R.L.~Disch.
\newblock {\em J. Am. Chem. Soc.}, 90:3104, 1968.

\bibitem{Buldakov1996}
M.A. Buldakov, I.I. Ippolitov, B.V. Korolev, I.I. Matrosov, A.E. Cheglokov,
  V.N. Chrerepanov, Yu.S. Makushkin, and O.N. Ulenikov.
\newblock {\em Spectrochim. Acta A}, 52:995, 1996.

\bibitem{Cohen1977}
E.~Cohen and Birnbaum G.J.
\newblock {\em J. Chem. Phys.}, 66:2443, 1977.

\bibitem{Bridge1966}
N.J. Bridge and A.D. Buckingham.
\newblock {\em Proc. R Soc. London Ser. A}, 295:334, 1966.

\bibitem{lepers2011a}
M.~Lepers, R.~Vexiau, N.~Bouloufa, O.~Dulieu, and V.~Kokoouline.
\newblock Photoassociation of a cold atom-molecule pair: second-order
  perturbation approach.
\newblock {\em Phys. Rev. A}, 83:042707, 2011.

\bibitem{lepers2011b}
M.~Lepers and O.~Dulieu.
\newblock Ultracold atom-dimer long-range interactions beyond the $1/r^{n}$
  expansion.
\newblock {\em Eur. Phys. J. D}, 165:113--123, 2011.

\bibitem{lepers2011c}
M.~Lepers and O.~Dulieu.
\newblock Long-range interactions between ultracold atoms and molecules
  including atomic spin-orbit.
\newblock {\em Phys. Chem. Chem. Phys.}, 13:19106--19113, 2011.

\bibitem{vdavoird1987}
A.~Van~der Avoird and G.~Brocks.
\newblock The {O}$_2$-{O}$_2$ dimer: magnetic coupling and spectrum.
\newblock {\em J. Chem. Phys}, 87:5346--5360, 1987.

\bibitem{dubernet1994}
M.-L. Dubernet and J.~M. Hutson.
\newblock Atom-molecule van der {W}aals complexes containing open-shell atoms.
  i. general theory and bending levels.
\newblock {\em J. Chem. Phys.}, 101:1939--1958, 1994.

\bibitem{chang1967}
T.~Y. Chang.
\newblock Moderately long-range interatomic forces.
\newblock {\em Rev. Mod. Phys.}, 39(4):911--942, 1967.

\bibitem{bishop1990}
D.~M. Bishop.
\newblock Molecular vibrational and rotational motion in static and dynamic
  electric fields.
\newblock {\em Rev. Mod. Phys.}, 62(2):343--374, 1990.

\bibitem{allison1986}
A.C. Allison, S.L. Guberman, and A.~Dalgarno.
\newblock A model of the {S}chumann-{R}unge continuum of {O$_2$}.
\newblock {\em J. Geophys. Res.}, 91(A9):10193--10, 1986.

\bibitem{friedman1990}
R.S. Friedman.
\newblock Oscillator strengths of the schumann-runge bands of isotopic oxygen
  molecules.
\newblock {\em J. Quant. Spec. Rad. Trans.}, 43:225--238, 1990.

\bibitem{wang1987}
J.~Wang, D.G. McCoy, A.J. Blake, and L.~Torop.
\newblock Effects of the close approach of potential curves in photoabsorption
  by diatomic molecules--ii. temperature dependence of the {O$_2$} cross
  section in the region 130-160 nm.
\newblock {\em J. Quant. Spec. Rad. Trans.}, 38:19--27, 1987.

\bibitem{singh1966}
R.B. Singh and D.K. Rai.
\newblock Potential-energy curves for {O}$_2^++$, {N}$_2^+$, and {CO}$^+$.
\newblock {\em J. Mol. Spec.}, 19:424--434, 1966.

\bibitem{spelsberg1998}
D.~Spelsberg and W.~Meyer.
\newblock Ab initio dynamic dipole polarizabilities for {O}$_2$, its
  photoabsorption spectrum in the schumann-runge region, and long-range
  interaction coefficients for its dimer.
\newblock {\em J. Chem. Phys.}, 109:9802--9810, 1998.

\bibitem{hathorn2000}
B.C. Hathorn and R.A. Marcus.
\newblock An intramolecular theory of the mass-independent isotope effect for
  ozone. ii. numerical implementation at low pressures using a loose transition
  state.
\newblock {\em J. Chem. Phys.}, 113:9497--9509, 2000.

\bibitem{avdeenkov2001}
A.~V. Avdeenkov and J.~L. Bohn.
\newblock Ultracold collisions of oxygen molecules.
\newblock {\em Phys. Rev. A}, 64:052703, 2001.

\bibitem{volpi2002}
A.~Volpi and J.L. Bohn.
\newblock Magnetic-field effects in ultracold molecular collisions.
\newblock {\em Phys. Rev. A}, 65:052712, 2002.

\bibitem{tilford2004}
K.~Tilford, M.~Hoster, P.M. Florian, and R.C. Forrey.
\newblock Cold collisions involving rotationally hot oxygen molecules.
\newblock {\em Phys. Rev. A}, 69:052705, 2004.

\bibitem{perez-rios2009}
J.~P\'erez-R\'ios, M.~Bartolomei, J.~Campos-Mart\'inez, M.I. Hern\'andez, and
  R.~Hern\'andez-Lamoneda.
\newblock Quantum-mechanical study of the collision dynamics of
  $\mathrm{O}_2(^3{\Sigma}_g^-) + \mathrm{O}_2(^3{\Sigma}_g^-)$ on a new ab
  initio potential energy surface.
\newblock {\em J. Phys. Chem. A}, 113:14952, 2009.

\bibitem{perez-rios2011}
J.~P{\'e}rez-R{\'\i}os, J.~Campos-Mart{\'\i}nez, and M.I. Hern{\'a}ndez.
\newblock Ultracold {O}$_2$ + {O}$_2$ collisions in a magnetic field: On the
  role of the potential energy surface.
\newblock {\em J. Chem. Phys.}, 134:124310, 2011.

\bibitem{patterson2007}
D.~Patterson and J.M. Doyle.
\newblock Bright, guided molecular beam with hydrodynamic enhancement.
\newblock {\em J. Chem. Phys.}, 126:154307, 2007.

\bibitem{narevicius2008}
E.~Narevicius, A.~Libson, C.G. Parthey, I.~Chavez, J.~Narevicius, U.~Even, and
  M.G. Raizen.
\newblock Stopping supersonic oxygen with a series of pulsed electromagnetic
  coils: A molecular coilgun.
\newblock {\em Phys. Rev. A}, 77:051401, 2008.

\bibitem{bucicov2008}
O.~Bucicov, M.~Nowak, S.~Jung, G.~Meijer, E.~Tiemann, and C.~Lisdat.
\newblock Cold {SO}$_2$ molecules by {S}tark deceleration.
\newblock {\em Eur. Phys. J. D}, 46:463, 2008.

\bibitem{reignier1998}
D.~Reignier, T.~Stoecklin, S.D. Le~Picard, A.~Canosa, and B.R. Rowe.
\newblock Rate constant calculations for atom--diatom reaction involving an
  open-shell atom and a molecule in a $\sigma$ electronic state application to
  the reaction {Al($^2P_{1/2,3/2}$)+O$_2(X^3\Sigma_g^-)\to$
  AlO($X^2\Sigma^+$)+O$(^3P_{2,1,0})$}.
\newblock {\em J. Chem. Soc., Faraday Trans.}, 94(12):1681--1686, 1998.

\bibitem{jorfi2010}
M.~Jorfi, B.~Bussery-Honvault, P.~Honvault, T.~Stoecklin, P.~Larr{\'e}garay,
  and P.~Halvick.
\newblock Theoretical sensitivity of the {C($^3P$)+OH($X^2\Pi$)$ \to
  $CO($X^1\Sigma^+$)+H($^2S$)} rate constant: The role of the long-range
  potential.
\newblock {\em J. Phys. Chem. A}, 114(28):7494--7499, 2010.

\bibitem{levine1987}
R.D. Levine and R.B. Bernstein.
\newblock {\em Molecular Reaction Dynamics and Chemical Reactivity}.
\newblock Oxford University Press, 1987.

\bibitem{clary1984}
D.C. Clary.
\newblock Rates of chemical reactions dominated by long-range intermolecular
  forces.
\newblock {\em Mol. Phys.}, 53(1):3--21, 1984.

\end{thebibliography}

\end{document}